\documentclass[onecolumn,11pt,aps,superscriptaddress]{revtex4-2}
\usepackage{gensymb}
\usepackage{notes2bib}
\usepackage{graphicx}
\usepackage{textcomp}
\usepackage{verbatim}
\usepackage{amsmath}
\usepackage{multirow}
\usepackage[dvipsnames]{xcolor}
\usepackage{subfigure}
\usepackage{tablefootnote}
\usepackage{wrapfig}
\usepackage{bm}
\usepackage{xfrac}
\usepackage{color}
\usepackage{xcolor}
\usepackage{setspace}
\usepackage{silence}
\usepackage{epstopdf}
\RequirePackage[normalem]{ulem}
\usepackage[colorlinks=true,urlcolor=blue,linkcolor=blue,citecolor=blue,a4paper]{hyperref}
\usepackage{soul}

\begin{document}
\title{Epitaxial Films and Devices of Transparent Conducting Oxides: La:BaSnO$_3$}

\author{Prosper Ngabonziza}
\email[corresponding author, ]{pngabonziza@lsu.edu}
\affiliation{Department of Physics and Astronomy, Louisiana State University, Baton Rouge,  LA 70803, USA}
\affiliation{Department of Physics, University of Johannesburg,P.O. Box 524 Auckland Park 2006, Johannesburg, South Africa}
\author{Arnaud P. Nono Tchiomo} 
\affiliation{Department of Physics and Astronomy, Louisiana State University, Baton Rouge,  LA 70803, USA}

\date{\today}

\begin{abstract}
This paper reviews recent developments in
materials science and device physics of high-quality epitaxial films of the transparent perovskite La-doped barium stannate, La:BaSnO$_3$. It presents current efforts in the synthesis science of epitaxial La:BaSnO$_3$ films for achieving reduced defect  densities and high electron mobility  at room temperature. We discuss the scattering mechanisms and the route towards engineering defect-free epitaxial La:BaSnO$_3$ heterostructures.  By combining chemical surface characterization and electronic transport studies, a special emphasis is laid on the proper correlation between the transport properties and the electronic band structure of La:BaSnO$_3$ films and heterostructures. For application purposes, interesting optical properties of La:BaSnO$_3$ films are discussed. Finally, for their potential application in oxide electronics, an overview of current progress in the fabrication of La:BaSnO$_3$-based thin-film field-effect transistors is presented together with recent progress in the the fundamental realization of two-dimensional electron gases with high electron mobility in La:BaSnO$_3$-based heterostructures. Future experimental studies to reveal the potential deployment of La:BaSnO$_3$ films in optoelectronic and transparent electronics are also discussed.
\end{abstract}

\pacs{???}


\maketitle
\newpage
\begin{center}
\textbf{1. Introduction}
\end{center} 
	
Research efforts in the field of materials science have led to discovering more complex and functional compounds that have greatly supported the advancement of cutting-edge technologies, especially in electronics. Hence, throughout the years, electronic devices have become light, robust, cheap, fast and more sustainable, and have made a huge impact to improving the living conditions of populations. An example of development in  materials science includes the synthesis/engineering of a new class of oxide materials that are transparent and conduct electricity with unconventional functionalities. These materials are referred to as transparent conducting oxides (TCOs). TCOs are materials that exhibit both high intermediate electrical conductivity ($\sim$10$^{-8}$ to 10$^3$ S cm$^{-1}$) and high optical transmission ($\geq$80\%). These materials are immensely used in the solar cell industry, and have revolutionized the screen display industry with transparent displays, organic light-emitting diodes (OLED) and liquid crystal displays (LCD)~\cite{kamiya2010present,ginley2000transparent,hHosono_2007,calnan2010high,hosono2002near}. Furthermore, TCOs have made devices smarter with the technology of touch panels, virtual reality and transparent-flexible electronics~\cite{nomura2004room,lee2017transparent}. However, due to the evolvement of new and competitive clean technologies that are based on perovskite oxides~\cite{Ngabonziza_2D_Doping_2021,Monama_2022,PNgabonziza_Inelastic_2021}, the demands for new TCOs that are particularly  eco-friendly, abundant and cheap to produce have increased, and it has sparked a surge of interest within the research community.
 
Oxidized thin metal films of cadmium (Cd) prepared in a glow discharge chamber were the first TCO ever reported, when B\"adeker demonstrated that these thin films could become transparent while remaining electrically conducting ~\cite{badeker1907concerning,lewis2000applications}. The capability of having two mutually exclusive properties such as current flow and transparency in a material is of high technological relevance, and has led to increasing research activities to develop and design new TCOs for specialized applications \cite{lewis2000applications,minami2000new,coutts2000characterization,freeman2000chemical}. Undoped and doped binary oxides such as In$_\text{2}$O$_\text{3}$, ZnO, SnO$_\text{2}$ and CdO have been considered as major \textit{n-}type TCO materials for several decades. This is primarily due to their remarkable process flexibility. They have an exceptional adhesion to most substrates and the possibility to be deposited at ambient temperature, which leads to various targeted applications. 
For example, in the design of optoelectronic devices, materials that exhibit maximum optical transmission in the visible spectrum coupled with high conductivity are preferred~\cite{ginley2000transparent,fortunato2007transparent,minami2000new,lewis2000applications,freeman2000chemical}. These TCOs exhibit degenerate semiconducting electrical characteristics, and their highly dispersive conduction band (CB) ensures plasma absorption in the infrared spectrum~\cite{ginley2000transparent,minami2000new,lewis2000applications}.

For industrial applications, the main principal markets include, but are not limited to, solar cell modules, flat-panel displays and architectural glass, where these TCOs are largely used. Important developments have been made in smart displays for the technology of smart devices (larger portable computers screens and larger flat-screen televisions of higher-resolution, touch screens and plasma displays) which are made from Sn-doped In$_\text{2}$O$_\text{3}$, as well as in the manufacturing of thin film solar cells, which are mostly made from Al-doped ZnO~\cite{ginley2000transparent,chopra1983transparent,kapur_1999,nakato1995improvement,green1976thin,heilmeier1970liquid,goodman1975liquid,muranoi1978properties,manifacier1981deposition,fortunato2007transparent}. Furthermore, thanks to the low production cost and the low emissivity of SnO$_\text{2}$, F-doped SnO$_\text{2}$ has been the material of choice in the production of functional glass. It is used as coated films in energy-efficient windows, whose transparency can be electrically controlled (electrochromic windows) and used for heat as well as light management in several domains such as aeronautic, automotive and housing~\cite{lewis2000applications,ginley2000transparent,chopra1983transparent,granqvist1998recent}. 
For application in electronic devices, TCOs are used as transparent conductive electrodes, transmitting light, and at the same time allowing the flow of electrical current. 

TCOs are processed following a variety of deposition methods, such as spraying onto hot glass substrates~\cite{coutts2000fundamental,aitchison1654transparent} and dc-magnetron sputtering in a mixture of Ar and O$_2$ background gas~\cite{holland1953properties,ow2005zito,coutts2000fundamental}. These preparation methods  lead to single crystalline thin films (the electrodes), and these films are not immune to crystallographic defects. The defects, mostly dislocations, stacking faults, point defects (oxygen vacancies) and grain boundaries, bring on additional scattering processes that contribute to damping out the electron mobility ($\mu_e$) in the films together with intrinsic ionized impurity scattering mechanisms~\cite{lewis2000applications,ellmer2012past}. As reported in  Ref.~\cite{lee2017transparent}, at a maximum electron doping level of 10$^{\text{21}}$~cm$^{-\text{3}}$, the room temperature electron mobility ($\mu_e^{\text{RT}}$) of these TCOs vary from $\mu_e^{\text{RT}}\approx\text{15}-\text{50}$~cm$^{\text{2}}$ V$^{-\text{1}}$~s$^{-\text{1}}$  for F-doped SnO$_\text{2}$, $\mu_e^{\text{RT}}\approx\text{20}-\text{100}$~cm$^{\text{2}}$ V$^{-\text{1}}$~s$^{-\text{1}}$  for Sn-doped In$_\text{2}$O$_\text{3}$, and $\mu_e^{\text{RT}}\approx\text{1}-\text{5}$~cm$^{\text{2}}$ V$^{-\text{1}}$~s$^{-\text{1}}$  for Al-doped ZnO~\cite{ellmer2012past,lee2017transparent}.

Although these binary TCO materials offer numerous processing advantages associated with their chemical composition, their relatively low carrier mobility is a limiting factor for integration in, for example, fast logic devices, where a high carrier mobility is a fundamental requirement for operation speed \cite{fortunato2008high,oana2001current}. Furthermore, the cost and scarcity of the indium (In) metal as well as the toxicity of Cd are other limitations that have triggered the search for new \textit{n-}type TCO compounds that would potentially feature superior physical properties~\cite{kumar2010race}. An archetype  of such compounds is the perovskite La-doped lead  zirconate titanate, La$_x$Pb$_{\text{1}-x}$(Zr$_y$Ti$_{\text{1}-y}$)O$_\text{3}$, also known as PLZT. PLZT has been employed as a solid-state material in several device applications including  photoconductor memory \cite{atkin1972performance}, switching devices \cite{maldonado1972ferroelectric}, infrared detectors \cite{liu1972pyroelectric}, as well as in signal processing \cite{kraut1972application}. In addition, PLZT has also been used in many optical applications such as display devices \cite{francombe1972research}, image sensors and storage \cite{schlosser1972self}, photochromics and electrochromics \cite{kiss1970photochromics}.

\begin{figure*}[!t]
  \centering
\includegraphics[width=0.75\textwidth]{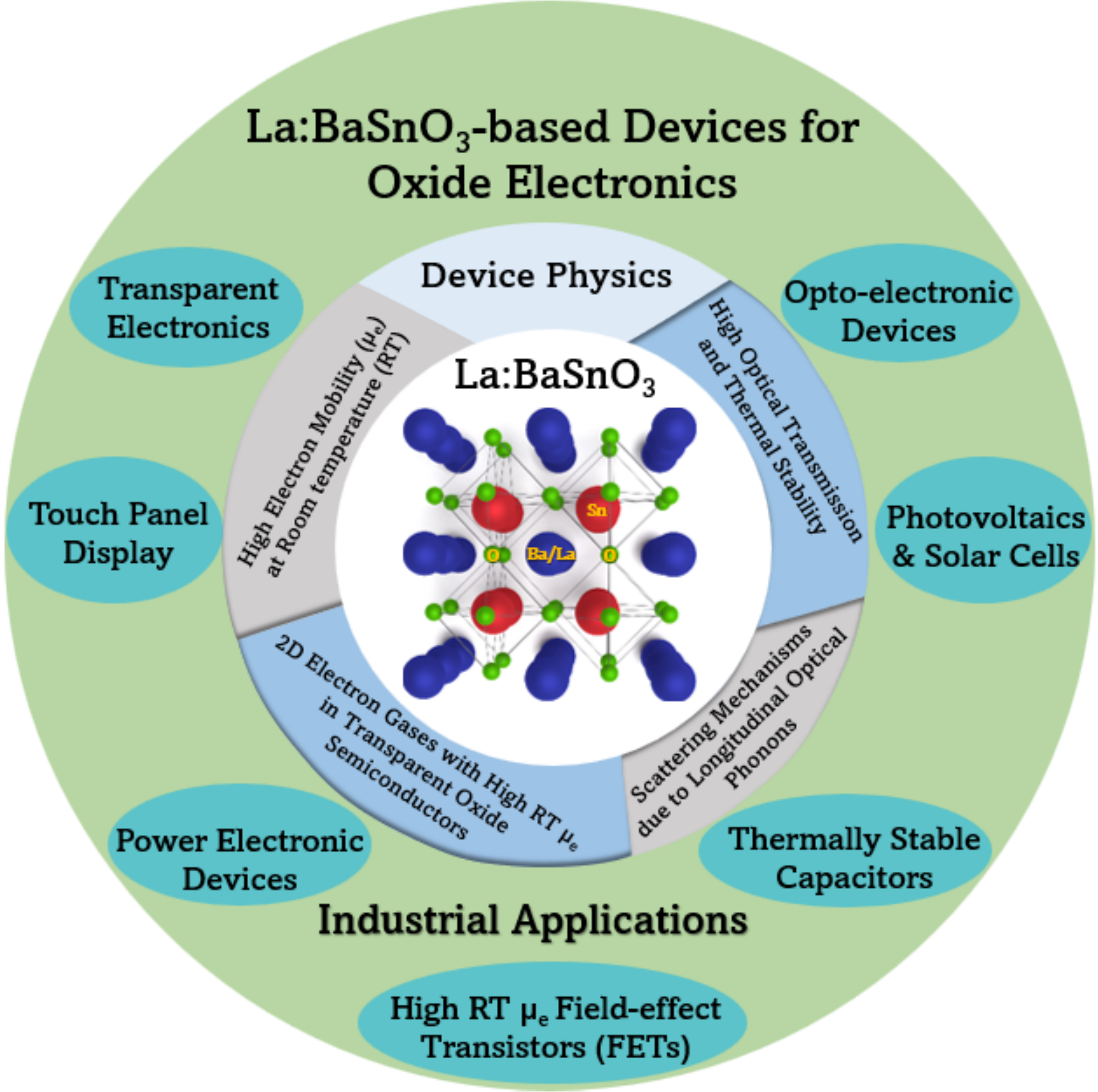}
  \caption{Schematic illustration of some key physical properties in epitaxial La:BaSnO$_3$ thin films. The high  room-temperature electron mobility ($\mu_e^{\text{RT}}$) with high optical transmission and thermal stability make La:BaSnO$_3$ an enticing material for the realization of emerging functionalities in thin films and heterostructures; also for the exploration of the physics of La:BaSnO$_3$-based devices and their potential practical applications in oxide electronics.}
  \label{fig:FermiSurface}
\end{figure*}

Very recently, a transparent perovskite oxide (barium tin oxide or barium stannate), with the chemical formula BaSnO$_3$, has been the subject of considerable attention within the scientific community due to its outstanding physical and electrical properties. When lightly doped, BaSnO$_3$ was found to display a degenerate \textit{n-}type semiconducting behavior marked by a high electrical conductivity while remaining optically transparent with a high optical transmission~\cite{kim2012high,kim2012physical,shimizu1985perovskite}. More importantly,  doped BaSnO$_3$ was found to exhibit an impressive reliability in the oxygen defect structure characterized by an exceptional stability in its electrical properties upon high temperature treatment in air atmosphere~\cite{kim2012high,kim2012physical}. These extraordinary performances suggest that electron doped BaSnO$_3$ is a TCO that has a great potential for application in oxide electronics. Not only that it could potentially be a solution for long-term operation of such devices in air, but also, it could be a great candidate for the replacement of Sn-doped In$_\text{2}$O$_\text{3}$ in the display industry due to its high $\mu_e^{\text{RT}}$~\cite{kim2012high}. Figure~\ref{fig:FermiSurface} depicts various device-physics utilization and a broad scope of technological applications of La-doped BaSnO$_3$-based materials. 

\begin{figure}[!t]
	\centering 
	\includegraphics[width=0.7\textwidth]{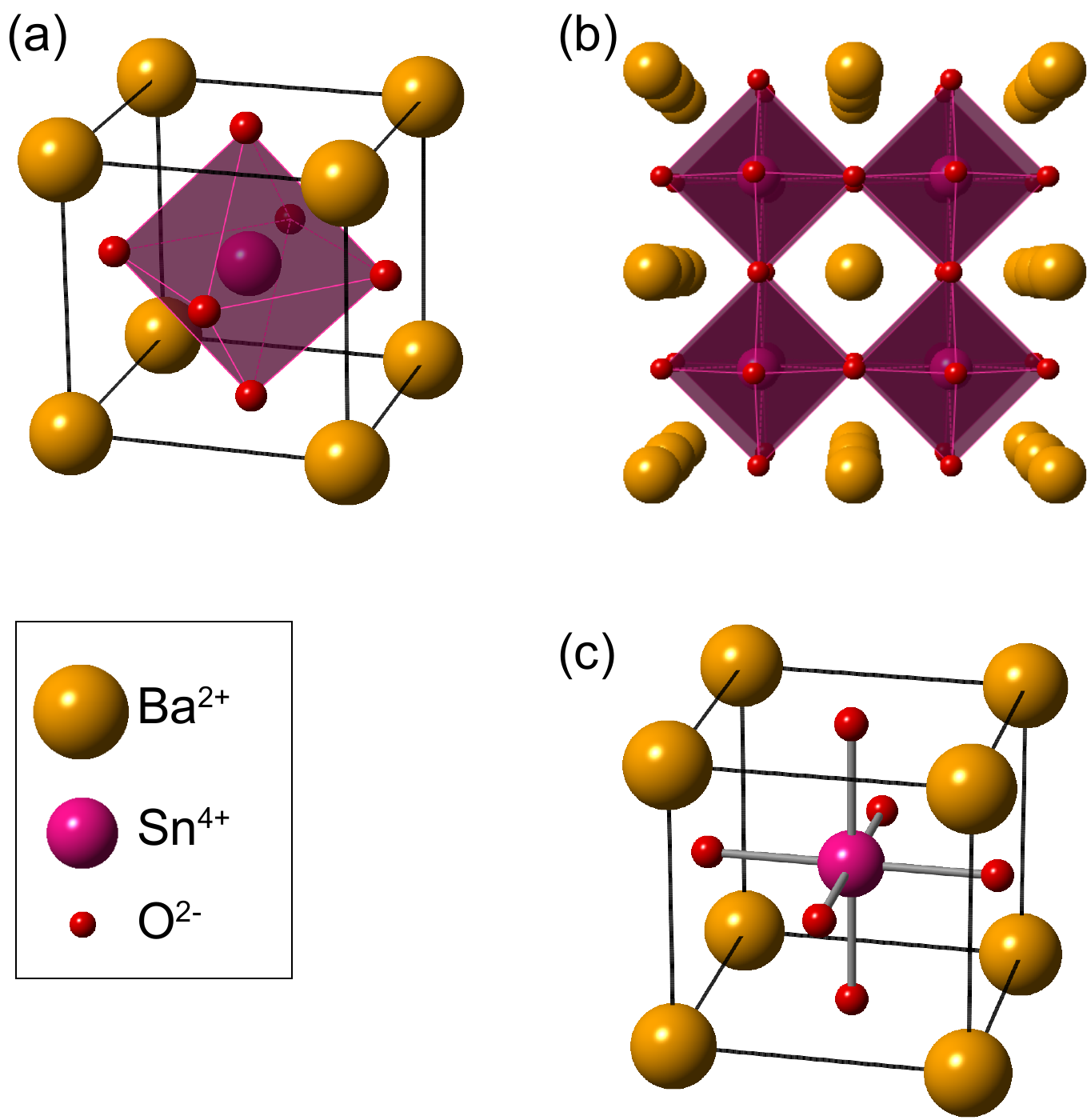}
	\caption{The perfect cubic structure of the transparent perovskite oxide BaSnO$_\text{3}$ (space group \textit{Pm\={3}m}). (a) A unit cell: the Ba$^{\text{2}+}$ ions occupy the corners of the cube; the Sn$^{\text{4}+}$ ions are at the center of the cube; and the $\text{O}^{\text{2}-}$ ions are located at the center of each face, giving a regular 3D octahedron around the Sn$^{\text{4}+}$. (b) No octahedral tilting distortion in the perovkite structure evidenced by a $\text{3}\times\text{3}\times\text{3}$ supercell, and (c) the perfect 180$\degree$ of O$^{\text{2}-}\text{---}\text{Sn}^{\text{4}+}\text{---}\text{O}^{\text{2}-}$ bonds. All the $\text{Sn}^{\text{4}+}\text{---}\text{O}^{\text{2}-}$ bonds are equal in length. This image is adapted from Ref.~\cite{tchiomo2020electronic}, Copyright \copyright\, 2020, Author, thesis distributed under a Creative Commons Attribution (CC BY) license.}
	\label{Crystal} 
\end{figure}

The  perovskite  material, BaSnO$_3$, has attracted much attention since 2012 after  Kim \textit{et al.} discovered that BaSnO$_3$ single crystal lightly doped with La (at a doping level of $\text{8}\times \text{10}^{\text{19}}$~cm$^{-\text{3}}$) exhibits an \textit{n-}type conductivity with a high $\mu_e^{\text{RT}}$ of 320~cm$^\text{2}$~V$^{-\text{1}}$~s$^{-\text{1}}$~ \cite{kim2012high}. These outstanding electrical characteristics were attributed to both the highly dispersive Sn~$5s$ orbitals derived CB of BaSnO$_3$, which is associated with a small electron effective mass ($m^*$)~ \cite{niedermeier2017electron,scanlon2013defect,mizoguchi2004probing}, and the low  scattering rate of optical phonons~\cite{prakash2017wide,sallis2013doped}. Figure~\ref{Crystal} depicts the perfect cubic perovskite structure of BaSnO$_{3}$. The crystal structure of the perovskite BaSnO$_3$ represents a key element in describing the origin of its electronic properties. Among the alkaline  earth  stannate (\textit{A}SnO$_\text{3}$, with \textit{A}=Ba, Sr and Ca), the material BaSnO$_{3}$ was found to have a simple cubic structure, characterized by a  network of corner-sharing SnO$_\text{6}$ octahedra, where the angles of the O---Sn---O bonds are exactly $\text{180}\degree$ (no octahedral tilting distortion)~\cite{wolfram2006electronic,tilley2016perovskites,mizoguchi2004probing,shannon1976revised,mountstevens2003cation}. 
 This network of SnO$_\text{6}$ octahedra is at the core of the conduction mechanism through the $5s$ and $2p$ states of the Sn$^{4+}$ and O$^{2-}$ ions, respectively; and it permits increased electron hopping between Sn$^{4+}$ ion sites~\cite{lee2017transparent,mizoguchi2004probing}. 

As compared to single crystal samples, the potential of La-doped BaSnO$_3$ (La:BaSnO$_{3}$)  for oxide electronic devices and for the fundamental realization of 2-dimensional electron gases with high $\mu_e^{\text{RT}}$  in transparent conducting oxide thin films and heterostructures has triggered considerable interests~\cite{PWilliam_2021,nono2022combined}. However, achieving a high electron mobility in La:BaSnO$_{3}$ thin films comparable to that obtained in the bulk single crystals has proved to be quite challenging. This is because these films contain a tremendous amount of dislocations defects, which in general result either from the growth conditions or from the large lattice mismatch with the substrate used in the epitaxial growth~\cite{wang2015atomic,sanchela2018large,yu2016enhancing,mun2013large,paik2017adsorption}. The main function of these dislocations is to operate as scattering cores and traps for the electrons, thereby reducing the number of free carrier and subsequently lowering the electron mobility~\cite{mun2013large}. 

Before the material La:BaSnO$_{3}$ could be implemented  as an active channel in electronic and optoelectronic devices [see, Fig.~\ref{fig:FermiSurface}], an electron mobility superior or equal to that achieved in traditional semiconductor systems is required. For several years, research activities have been carried out to improve the quality and electronic performances of epitaxial La:BaSnO$_{3}$ films. 

This paper reviews recent progress in materials science and device physics of epitaxial films of La:BaSnO$_3$. Recent advances in thin-film growth for achieving reduced defect densities and high $\mu_e^{\text{RT}}$ in La:BaSnO$_{3}$ films are discussed. A detailed discussion on the scattering mechanisms and potential routes towards engineering defect-free epitaxial La:BaSnO$_3$ heterostructures are presented.  By combining chemical surface characterization and electronic transport studies, a special emphasis is laid on the proper correlation between the transport properties and the electronic band structure of La:BaSnO$_3$ heterostructures. Furthermore, for implementation of a variety of La:BaSnO$_3$-based electronic devices,  an overview of current progress in the fabrication of thin-film field-effect transistors is presented. Finally, we discuss future directions that will focus on exploring the physics of La:BaSnO$_3$ heterostructures for their applications in oxide electronic and optoelectoronic devices.

\begin{center}
\textbf{2. Synthesis science and electronic transport properties of La:BaSnO$_3$ films}
\end{center}

In order to enhance the quality of epitaxial La:BaSnO$_{3}$ films and achieve improvement in the electronic transport properties, several synthesis approaches have been employed. This review focuses on the most widely used deposition techniques, namely molecular beam epitaxy (MBE) and pulsed laser deposition (PLD). Other deposition techniques that have been used for the deposition of La:BaSnO$_{3}$ epitaxial films include, but are not limited to, high-pressure magnetron sputtering~\cite{RZhang_2021}, high-pressure-oxygen sputter deposition~\cite{PWilliam_2021} and chemical solution deposition technique~\cite{YHe_2021}.

MBE and/or PLD thin film deposition methods have been used in the fabrication of epitaxial La:BaSnO$_{3}$ films to achieve high crystal quality. The reported processing approaches firstly include the use of lattice matched substrates~\cite{lee2016enhanced}; secondly, the use of an insulating buffer layer that is inserted between the active film layer and the substrate to minimize or annihilate the interfacial defects originating from lattice mismatches \cite{paik2017adsorption,niedermeier2017electron,shiogai2016improvement,prakash2017wide,wang2019epitaxial}; and thirdly, the application of post-growth treatments to improve the film quality by annealing the as-grown films in vacuum or in gas environment~\cite{yu2016enhancing,yoon2018oxygen,cho2019effects}. Here, we indicate that in the growth process, perovskite oxide single crystalline substrates which offer both lattice and structural matches have been preferably used ~\cite{shin2016high,niedermeier2017electron,shiogai2016improvement,paik2017adsorption,wang2019epitaxial}.

\begin{figure}[!t]
	\centering 
	\includegraphics[width=1\textwidth]{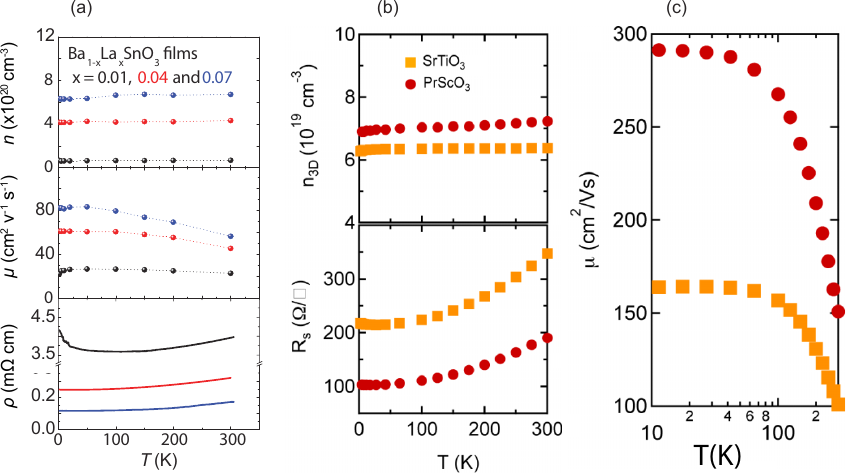}
	\caption{Temperature dependence of the electronic transport characteristics, namely, the  carrier density (\textit{n} and n$_{\text{3D}}$), the electron mobility ($\mu$), and the resistivity ($\rho$ and R$_\text{s}$) for La:BaSnO$_{3}$ thin films grown by (a) PLD on SrTiO$_\text{3}$ (001), and (b) and (c) MBE on PrScO$_\text{3}$ (110) and SrTiO$_\text{3}$ (001) substrates. The image in (a) is reprinted with permission from Ref.~\citep{kim2012physical}, Copyright\,\copyright\,2012 American Physical Society. The images in (b) and (c) are reprinted and adapted with permission from Ref~\cite{raghavan2016high}, Copyright \copyright\, 2016, Author(s) under a Creative Commons Attribution (CC BY) license.}
	\label{Kim2012a-Raghavan2016a} 
\end{figure} 

Early in the synthesis of epitaxial La:BaSnO$_{3}$ thin films, remarkable achievements in growing high mobility thin films have been reported~\cite{kim2012high,kim2012physical,mun2013large}. Epitaxial thin films containing 1\%, 4\% and 7\%  La doping were deposited directly on SrTiO$_\text{3}$ (001) substrates using PLD, and a $\mu_e^{\text{RT}}$ up to $\text{70}$~cm$^\text{2}$~V$^{-\text{1}}$~s$^{-\text{1}}$ with a carrier concentration, $n$, of  $~\approx\text{6}\times \text{10}^{\text{20}}$~cm$^{-\text{3}}$ were reported in the 7\% La-doped films.  A few years later, MBE-grown films of La:BaSnO$_{3}$ were directly deposited on SrTiO$_\text{3}$ (001) substrates, and an improved $\mu_e^{\text{RT}}\approx\text{124}$~cm$^\text{2}$~V$^{-\text{1}}$~s$^{-\text{1}}$ at  $\textit{n}\approx\text{6.4}\times \text{10}^{\text{19}}$~cm$^{-\text{3}}$  was achieved~\cite{raghavan2016high}. Figure~\ref{Kim2012a-Raghavan2016a} depicts the temperature dependence of the electronic transport properties of these films, including the carrier densities \textit{n} and n$_{\text{3D}}$, the electron mobilities $\mu$, and the resistivities $\rho$ and R$_\text{s}$. The measured \textit{n} and n$_{\text{3D}}$ are almost temperature independent [see, upper panels of Fig.~\ref{Kim2012a-Raghavan2016a}\textcolor{blue}{(a)} and Fig.~\ref{Kim2012a-Raghavan2016a}\textcolor{blue}{(b)} (orangish markers)].  This behavior, associated with the metallic characteristic of $\rho$ and R$_\text{s}$ as illustrated in the lower panels [see, Fig.~\ref{Kim2012a-Raghavan2016a}\textcolor{blue}{(a)} and Fig.~\ref{Kim2012a-Raghavan2016a}\textcolor{blue}{(b)}], is indicative of a degenerate semiconducting regime for all these films~\cite{kim2012physical,raghavan2016high}. The $\mu$ values decrease with increasing temperature [see,  Fig.~\ref{Kim2012a-Raghavan2016a}\textcolor{blue}{(a)} (middle panel) and Fig.~\ref{Kim2012a-Raghavan2016a}\textcolor{blue}{(c)}]. Note that in Fig.~\ref{Kim2012a-Raghavan2016a}\textcolor{blue}{(a)}, $\mu$ decreases from  2~K up to 300~K~\cite{kim2012physical}, but in Fig.~\ref{Kim2012a-Raghavan2016a}\textcolor{blue}{(c)}, there is first a saturation of  $\mu$ up to about 100~K, and then the decrease follows~\cite{raghavan2016high}. For several other reported epitaxial La:BaSnO$_{3}$ films grown by MBE and PLD, similar temperature dependence of the electrical transport properties (\textit{n}, $\rho$ and $\mu$) were observed~\cite{prakash2017wide,prakash2017adsorption,niedermeier2016solid,shiogai2016improvement,wadekar2014improved,paik2017adsorption,wang2019epitaxial}.

\begin{figure}[!t]
	\centering 
	\includegraphics[width=1\textwidth]{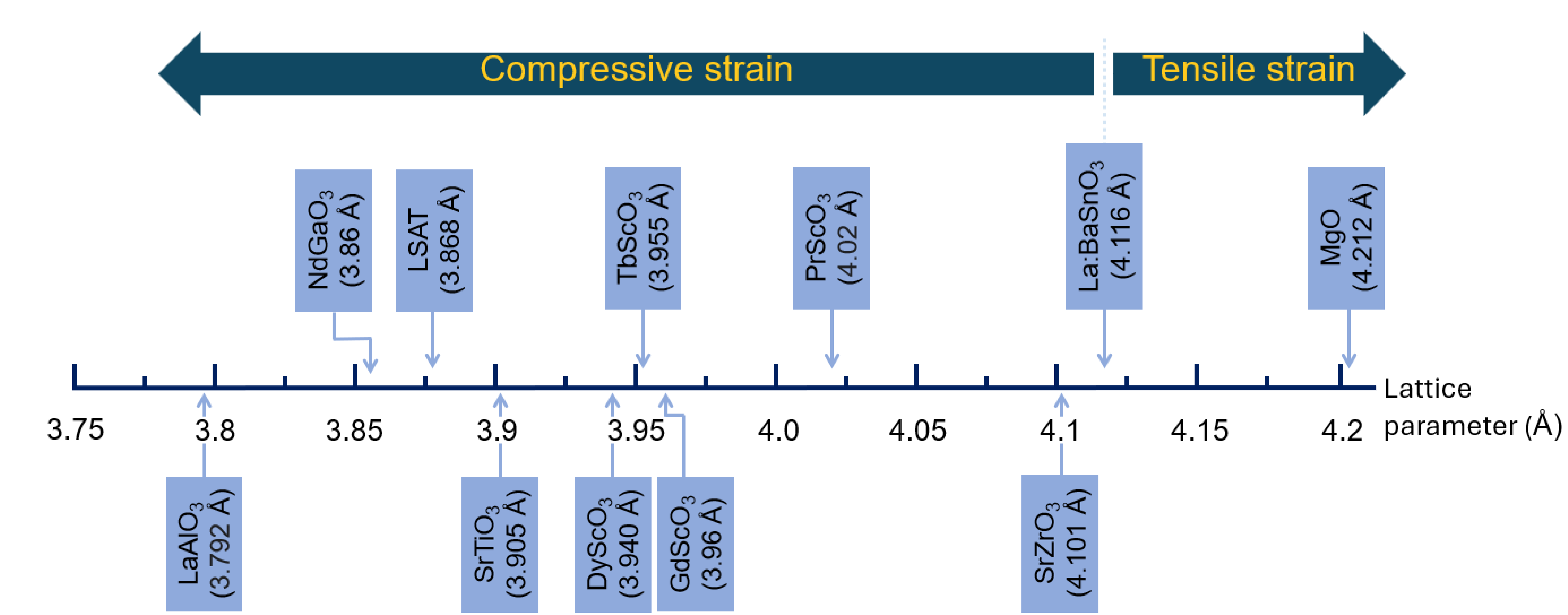}
	\caption{Diagram presenting the lattice parameters of some commercially available oxide single crystal substrates, and compared to that of La:BaSnO$_{3}$ material. Note that SrZrO$_3$ is not commercially available as a single crystal substrate, and its relevance in this diagram will be discussed later on in the text. Reprinted and adapted with permission from  Ref.~\cite{ngabonziza2023employing}, Copyright \copyright\, 2023 AIP Publishing.}
	\label{commercilaLAttice} 
\end{figure}

It is known that the flow of electrons in epitaxial La:BaSnO$_{3}$ films is hindered by the presence of different types of scattering sources, the most prominent being the threading dislocations (TDs). In these films, the TDs originate from the large lattice mismatch between the as-grown film and the substrate. The reduction of the  density  of these TDs  has been demonstrated to significantly  help in the improvement of the electron mobility in La:BaSnO$_{3}$ epitaxial films~\cite{ngabonziza2023employing,nono2019high}. These mobility limiting factors will be discussed in more detail in the next section. 

The epitaxial growth of high-quality La:BaSnO$_{3}$ films free from these structural defects would require the use of a substrate material with a lattice parameter close to the film's lattice constant. BaSnO$_\text{3}$ is therefore the suitable candidate material to use in the epitaxial growth. By using the PLD, BaSnO$_\text{3}$ has been used as substrate material by Lee \textit{et al.} \cite{lee2016enhanced}.  Epitaxial thin films of La$_x$Ba$_{\text{1}-x}$SnO$_\text{3}$ (with several La doping content, $x$, varying from 0.005 to 0.04) were successfully deposited directly on BaSnO$_\text{3}$ single crystal  substrates. Although their microstructural analysis did not reveal any evidence of defects (dislocations or grain boundaries) at the interface between the films and substrate, as well as within the film area, only a maximum  $\mu_e^{RT}$ of 100~cm$^\text{2}$~V$^{-\text{1}}$~s$^{-\text{1}}$ at $\textit{n}=\text{1.3}\times \text{10}^{\text{20}}$~cm$^{-\text{3}}$ was achieved. 

Ideally, employing BaSnO$_\text{3}$ as substrate would be the best option for achieving high electron mobility in the active layer of La:BaSnO$_{3}$ films grown on top. But, as the synthesis science of BaSnO$_\text{3}$ single crystalline has not yet been fully mastered, it is still difficult to produce these single crystalline substrates.  Thus,  it is not commercially available, and researchers had to develop other alternatives like to use other substrates with closest lattice parameters to that of BaSnO$_\text{3}$ material.

\begin{table*}[!b]
	\begin{center}
		\caption{Synopsis of the synthesis of high electron mobility La:BaSnO$_\text{3}$ thin films.}
		\label{Synopsis-Progress}
		\setlength{\tabcolsep}{6pt}
		\begin{tabular}{cccccc}
			\hline
			\hline
			\small{Authors} & \small{Growth}  & \small{Substrate} & \small{Buffer layer}  & \small{RT mobility} &  \small{RT density} \\
			\small{} & \small{method} & \small{} & \small{} & \small{(cm$^\text{2}$~V$^{-\text{1}}$~s$^{-\text{1}}$)} & \small{(cm$^{-\text{3}}$)} \\
			\multicolumn{1}{l}{\small{Sallis \textit{et al.} \cite{sallis2013doped}}} & \small{PLD} & \small{SrTiO$_\text{3}$ (001)} & \small{---} & \small{48.5} & \small{2.2$\times \text{10}^{\text{20}}$}  \\
			\multicolumn{1}{l}{\small{Kim \textit{et al.} \cite{kim2012physical}}} & \small{PLD} & \small{SrTiO$_\text{3}$ (001)} & \small{---} & \small{70} & \small{$\approx$~6$\times \text{10}^{\text{20}}$}  \\
			\multicolumn{1}{l}{\small{Prakash \textit{et al.} \cite{prakash2017adsorption}}} & \small{MBE} & \small{SrTiO$_\text{3}$ (001)} & \small{---} & \small{105} &  \small{2.5$\times \text{10}^{\text{20}}$} \\
			\multicolumn{1}{l}{\small{Raghavan \textit{et al.} \cite{raghavan2016high}}} & \small{MBE} & \small{SrTiO$_\text{3}$ (001)} & \small{---} & \small{$\approx$~124} &  \small{$\approx$~6$\times \text{10}^{\text{19}}$} \\
			\multicolumn{1}{l}{\small{Lee \textit{et al.} \cite{lee2016enhanced}}} & \small{PLD} & \small{BaSnO$_\text{3}$ (001)} & \small{---} & \small{100} & \small{1.3$\times \text{10}^{\text{20}}$}  \\
			\multicolumn{1}{l}{\small{Lebens \textit{et al.} \cite{lebens2016direct}}} & \small{MBE} & \small{TbScO$_\text{3}$ (110)} & \small{---} & \small{81} & \small{1.65$\times \text{10}^{\text{20}}$}  \\
			\multicolumn{1}{l}{\small{Raghavan \textit{et al.} \cite{raghavan2016high}}} & \small{MBE} & \small{PrScO$_\text{3}$ (110)} & \small{---} & \small{150} & \small{$\approx$~7$\times \text{10}^{\text{19}}$}  \\
			\multicolumn{1}{l}{\small{Park \textit{et al.} \cite{park2014high}}} & \small{PLD} & \small{SrTiO$_\text{3}$ (001)} & \small{BaSnO$_\text{3}$} & \small{66} & \small{1$\times \text{10}^{\text{20}}$}  \\
			\multicolumn{1}{l}{\small{Niedermeier \textit{et al.} \cite{niedermeier2017electron}}} & \small{PLD} & \small{MgO} & \small{NiO} & \small{70} & \small{4.4$\times \text{10}^{\text{20}}$}  \\
			\multicolumn{1}{l}{\small{Shiogai \textit{et al.} \cite{shiogai2016improvement}}} & \small{PLD} & \small{SrTiO$_\text{3}$ (001)} & \small{BaSnO$_\text{3}$} & \small{78} & \small{8.5$\times \text{10}^{\text{19}}$}  \\
			\multicolumn{1}{l}{\small{Shin \textit{et al.} \cite{shin2016high}}} & \small{PLD} & \small{MgO} & \small{BaSnO$_\text{3}$} & \small{97.2} & \small{2.53$\times \text{10}^{\text{20}}$}  \\
			\multicolumn{1}{l}{\small{Prakash \textit{et al.} \cite{prakash2017wide}}} & \small{MBE} & \small{SrTiO$_\text{3}$ (001)} & \small{BaSnO$_\text{3}$} & \small{120} & \small{3$\times \text{10}^{\text{20}}$}  \\
			\multicolumn{1}{l}{\small{Wang \textit{et al.} \cite{wang2019epitaxial}}} & \small{MBE} & \small{Si (001)} & \small{BaSnO$_\text{3}$} & \small{128} & \small{1.4$\times \text{10}^{\text{20}}$}  \\
			\multicolumn{1}{l}{\small{Paik \textit{et al.} \cite{paik2017adsorption}}} & \small{MBE} & \small{DyScO$_\text{3}$ (001)} & \small{BaSnO$_\text{3}$} & \small{183} & \small{1.2$\times \text{10}^{\text{20}}$}  \\
			\hline
			\hline
		\end{tabular}
	\end{center} 
\end{table*}

Figure~\ref{commercilaLAttice} depicts a diagram of lattice constants for some commercially available oxide single crystalline substrates. The one with the closest lattice match is the perovskite scandate PrScO$_\text{3}$. There is a lattice mismatch of only $-\text{2.18}$\% (compressive strain) between the BaSnO$_\text{3}$ film and PrScO$_\text{3}$ (110) substrate~\cite{raghavan2016high}. This lattice mismatch is smaller compared to the  $-\text{5.4}$\%  mismatch between BaSnO$_\text{3}$ material and SrTiO$_\text{3}$ (001) substrates. By using MBE, an enhanced $\mu_e^{RT}$ up to 150~cm$^\text{2}$~V$^{-\text{1}}$~s$^{-\text{1}}$ with $\textit{n}\approx$~7.1$\times \text{10}^{\text{19}}$~cm$^{-\text{3}}$ was reported for La:BaSnO$_{3}$ films grown directly on PrScO$_\text{3}$ (110) substrates   [see plot with red markers in Fig.~\ref{Kim2012a-Raghavan2016a}\textcolor{blue}{(b)} and Fig.~\ref{Kim2012a-Raghavan2016a}\textcolor{blue}{(c)}] \cite{raghavan2016high}. The difference in lattice mismatch explains the difference in mobility values recorded in La:BaSnO$_\text{3}$ films prepared on PrScO$_\text{3}$ (110) and SrTiO$_\text{3}$ (001) substrates [see, Fig.~\ref{Kim2012a-Raghavan2016a}\textcolor{blue}{(c)}].

An alternative approach, which has also been commonly used in the epitaxial growth of La:BaSnO$_\text{3}$ films, is to insert an insulating layer, known as buffer layer, between the active layer of the film and the substrate. BaSnO$_\text{3}$ material has been often used as a buffer layer in the growth of La:BaSnO$_\text{3}$ thin films. The principal function of the buffer layer is to reduce the large lattice mismatch between the active film layer and the underlying substrate. Depositing thick buffer layers has been demonstrated to reduce the density of dislocations because  these dislocations are prone to disappear in the middle of the epitaxial growth of these thick buffer layer films~\cite{jasinski2002extended,farahani2012influence}. From available data in literature, in order to reach high electron mobilities,  the optimized thickness of the BaSnO$_\text{3}$ buffer layer was found to vary between 50 to 330~nm~\cite{park2014high,niedermeier2017electron,shiogai2016improvement,shin2016high,prakash2017wide,wang2019epitaxial,paik2017adsorption}. 

Table~\ref{Synopsis-Progress} gives a synopsis for the efforts made in synthesizing high electron mobility unbuffered and buffered La:BaSnO$_{3}$ thin films.  For the films grown by MBE on DyScO$_\text{3}$ (001) substrates and  containing a 330~nm BaSnO$_\text{3}$ buffer layer, a record $\mu^{\text{RT}}$ of 183~cm$^\text{2}$~V$^{-\text{1}}$~s$^{-\text{1}}$ at  $\textit{n}=\text{1.2}\times \text{10}^{\text{20}}$~cm$^{-\text{3}}$ was achieved~\cite{paik2017adsorption}. Considering all epitaxial growth techniques, this value is the highest reported $\mu^{\text{RT}}$ in La:BaSnO$_{3}$-based thin films to date. The results in Table~\ref{Synopsis-Progress} point to the notable enhancement of the electron mobility with the use of both scandate substrates and buffer layers. All the techniques developed above to synthesis La:BaSnO$_{3}$ thin films with improved electron mobility suggest that there is room to develop new strategies to eliminate most of the source of electron scattering in these epitaxial films so that the electron mobility could be boosted further.

\begin{center} 
\textbf{3. Scattering mechanisms and route for engineering dislocations-free epitaxial La:BaSnO$_3$ heterostructures}
\end{center}

\begin{center} 
\textit{3.1. Scattering mechanisms and role of threading dislocations for limiting $\mu_e$ in La:BaSnO$_3$}
\end{center}

Scattering mechanisms are the main limitation source of electron mobility in epitaxially grown La:BaSnO$_3$ thin films \cite{paik2017adsorption,raghavan2016high,niedermeier2016solid,wang2015atomic,shiogai2016improvement,kim2014dopant}. For example, in \textit{n-}type Si-doped GaN films, it has been demonstrated that at low doping concentrations   $\leq 5\times\text{10}^{\text{17}}$~cm$^{-\text{3}}$, scattering by dislocations is the leading mechanism; whereas at higher doping regimes, it is the scattering by ionized impurity that is dominant~\cite{weimann1998scattering,ng1998role,look1999dislocation}. However, it should be noted that in degenerate semiconductor systems, lattice scattering mechanisms which include phonon scattering, scattering at stacking faults and point defects are also at play, and should be considered~\cite{weimann1998scattering,ng1998role}. Hence, analytical, theoretical, and experimental studies have been largely carried out to investigate the contribution of these scattering mechanisms to the electronic transport properties of epitaxial films ~\cite{blatt1968physics,kasap2007springer,nag1972theory,look2001dislocation,niedermeier2017electron,krishnaswamy2017first,scanlon2013defect,prakash2017wide}. 

Analytically, it has been reported that for epitaxial thin films in degenerately doped semiconducting regime, the total electron mobility, $\mu_{\text{tot}}$, is governed by the scattering mechanisms described  above. The total electron mobility can  be modelled using Matthiesen's rule as \cite{blatt1968physics,kasap2007springer,nag1972theory}
\begin{align}
\label{Chap5-eqat1}
\frac{1}{\mu_{\text{tot}}} &=\frac{1}{\mu_{\text{dis}}}+\frac{1}{\mu_{\text{Imp}}}+\frac{1}{\mu_{\text{latt}}}.
\end{align}
where $\mu_{\text{dis}}$, $\mu_{\text{Imp}}$, and $\mu_{\text{latt}}$ are the mobility contributions due to dislocation, impurity and lattice scattering, respectively. For an elaborated introduction on the Matthiesen's rule, especially detailed discussion of the individual components on the right hand side, the interested reader can consult available references on this topic~\cite{look2001dislocation,niedermeier2017electron,low1955mobility,frohlich1939mean,thongnum2023description}. 
These components show a strong dependence of scattering by dislocation and scattering by impurity on the net carrier density, thereby highlighting their aforementioned  contribution to the overall mobility with respect to the doping level. 

These analytical models were also reported to describe the mobility limiting mechanisms in  \textit{n-}type La:BaSnO$_\text{3}$ systems. The degenerate regime was found to be reached at doping levels exceeding $\text{1}\times\text{10}^{\text{19}}$~cm$^{-\text{3}}$ ~\cite{prakash2017wide,niedermeier2017electron,kim2012physical,kim2013indications}. It was also reported that in  degenerate \textit{n-}type La:BaSnO$_\text{3}$ systems, the density of ionized impurity does not depend on the temperature, and as a result, ionized impurity scattering governs the low temperature transport properties \cite{look1997degenerate,look1997defect}. In literature, the scattering by phonon has been considered negligible at low temperature, but it acts as a significant factor of electron mobility limiting mechanism at high temperature, together with dislocation and impurity scattering \cite{prakash2017wide,kim2012physical,look2001dislocation,thongnum2023description}.

\begin{figure}[!t]
	\centering 
	\includegraphics[width=1\textwidth]{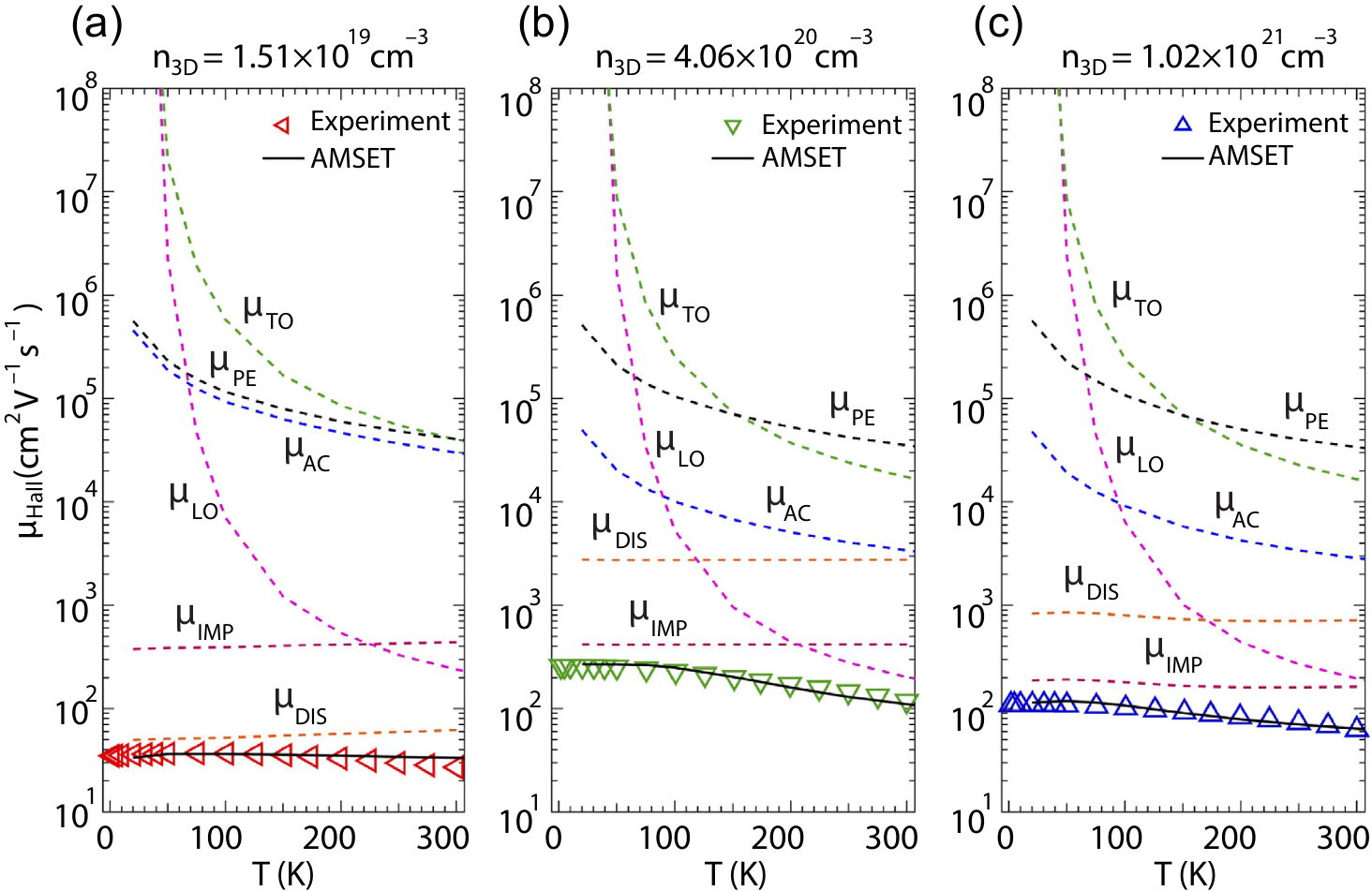}
	\caption[FootNote1]{(a)-(c) Temperature dependence of electron mobility ($\mu$) for La-doped BaSnO$_\text{3}$ samples at different La doping levels, with the calculated contribution of several scattering mechanisms (dashed lines) such as PE ($\mu_{\text{PE}}$), DIS ($\mu_{\text{DIS}}$), AC ($\mu_{\text{AC}}$), LO ($\mu_{\text{LO}}$), TO ($\mu_{\text{TO}}$) and IMP ($\mu_{\text{IMP}}$). The experimental data (symbols) are plotted alongside the calculated values (solid lines) obtained from DFT employing the AMSET  package. These images  were adapted from Ref.~\cite{prakash2017wide} with copyright \copyright\, 2017, Author(s) under a Creative Commons Attribution (CC BY) license. }
	\label{Prakash2017} 
\end{figure}

There are also numerous reports of DFT calculations simulating the effect of these scattering mechanisms on the transport properties of La:BaSnO$_3$ systems~\cite{niedermeier2017electron,krishnaswamy2017first,scanlon2013defect,prakash2017wide}. For instance, Prakash \textit{et al.} \cite{prakash2017wide}  conducted some works by analyzing the  temperature dependence of the mobilities for as-grown La:BaSnO$_\text{3}$ thin films and calculated bulk BaSnO$_\text{3}$, where the mobility contributions of scattering mechanisms such as, piezoelectric (PE), acoustic phonon deformation potential (AC), longitudinal polar optical phonon (LO), transverse optical phonon (TO), threading dislocations (DIS), and ionized impurities (IMP) were included.  Figure~\ref{Prakash2017} depicts the temperature dependence of $\mu$ for La-doped BaSnO$_\text{3}$ samples at different La doping levels together with the calculated contribution of several scattering mechanisms. The results of these studies assume various doping regimes that include, for example, the low ($\text{1.51}\times \text{10}^{\text{19}}$~cm$^{-\text{3}}$), intermediate ($\text{4.06}\times \text{10}^{\text{20}}$~cm$^{-\text{3}}$) and high ($\text{1.02}\times \text{10}^{\text{21}}$~cm$^{-\text{3}}$) regimes.  

Focusing our attention on the most dominant mechanisms (DIS and IMP), their  mobility limiting effects are clearly evidenced in the entire temperature range, especially in the  low [see, Fig.~\ref{Prakash2017}\textcolor{blue}{(a)}] and the high [see, Fig.~\ref{Prakash2017}\textcolor{blue}{(c)}] doping levels. In the intermediate doping regime, IMP scattering also has a dominating effect, but only below about 200~K [see, Fig.~\ref{Prakash2017}\textcolor{blue}{(b)}]. The behaviour of $\mu_{\text{DIS}}$ with respect to the  electron density is associated with the calculated values of both the scattering rate, which  was found to be larger for the low doping and lower for the other two regimes, and the dislocation density, which from low to high doping are $\text{2.3}\times \text{10}^{\text{12}}$~cm$^{-\text{2}}$, $\text{7.0}\times \text{10}^{\text{11}}$~cm$^{-\text{2}}$ and $\text{2.8}\times \text{10}^{\text{12}}$~cm$^{-\text{2}}$, respectively. This suggests that the contribution of DIS to the mobility would increase from low to intermediate regime, before decreasing again at high doping due to the combined effect of dislocation density and  inverse screening length (which is smaller at low doping level).

\begin{figure}[!t]
	\centering 
	\includegraphics[width=0.7\textwidth]{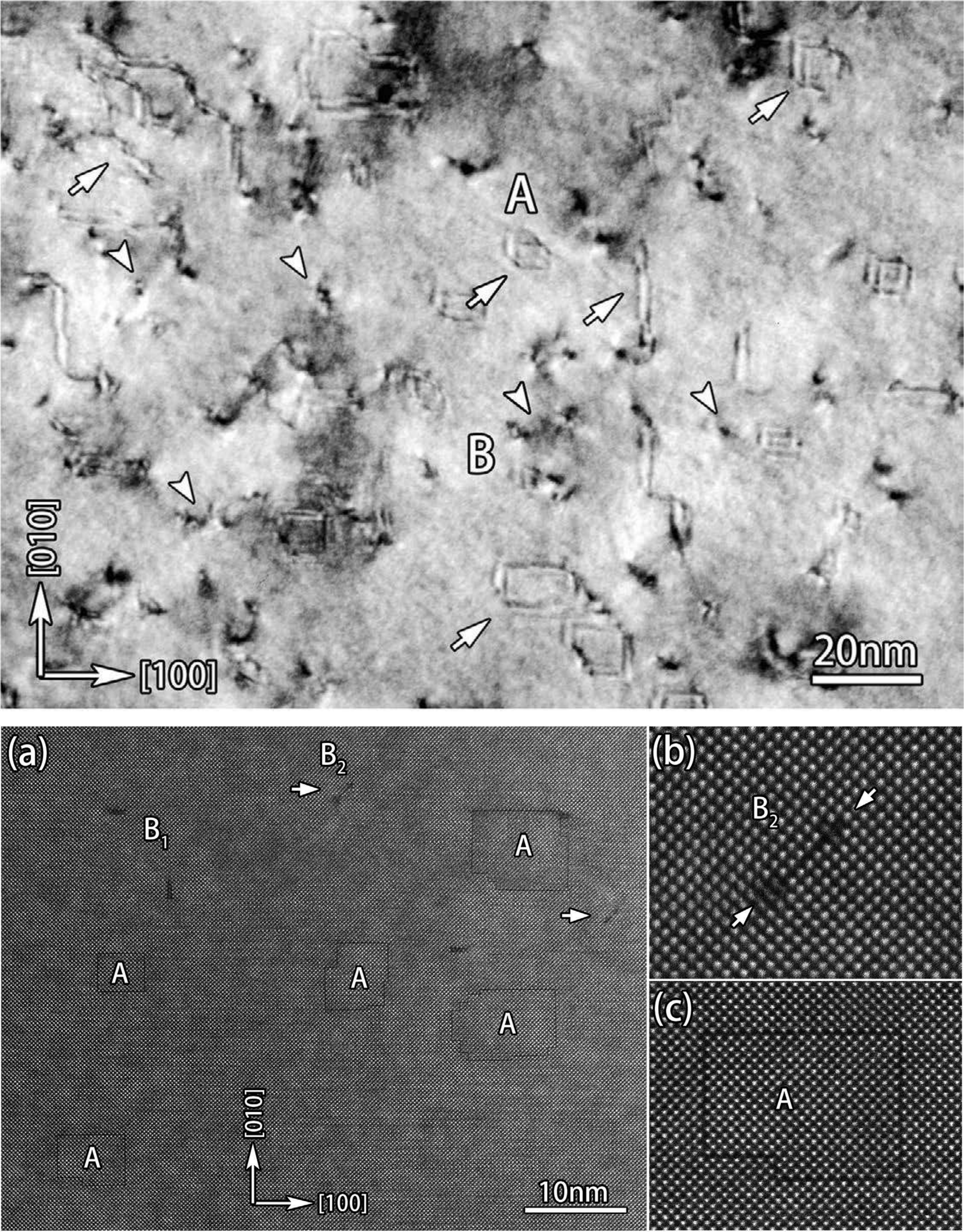}
	\caption{Plan-view STEM images of a 0.001\% La-BaSnO$_\text{3}$/LaScO$_\text{3}$/SrTiO$_\text{3}$ film. Upper panel: bright field image showing stacking faults indicated with long arrows (line contrasts A), and threading dislocations marked with short arrows (dark spots B). Lower panel: low magnification high angle annular dark field (LM-HAADF) STEM images presenting (a) polygon-like shapes stacking faults denoted by A, and two types of threading dislocations (B$_\text{1}$ and B$_\text{2}$). The defects B$_\text{2}$ and A are atom resolved and shown in (b) and (c), respectively. The images were adapted from Ref.~\cite{wang2015atomic} Copyright \copyright\, 2015, the Author(s) under a Creative Commons Attribution (CC BY) license.} 
	\label{Wang2015} 
\end{figure}

\begin{figure}[!t]
	\centering 
	\includegraphics[width=0.9\textwidth]{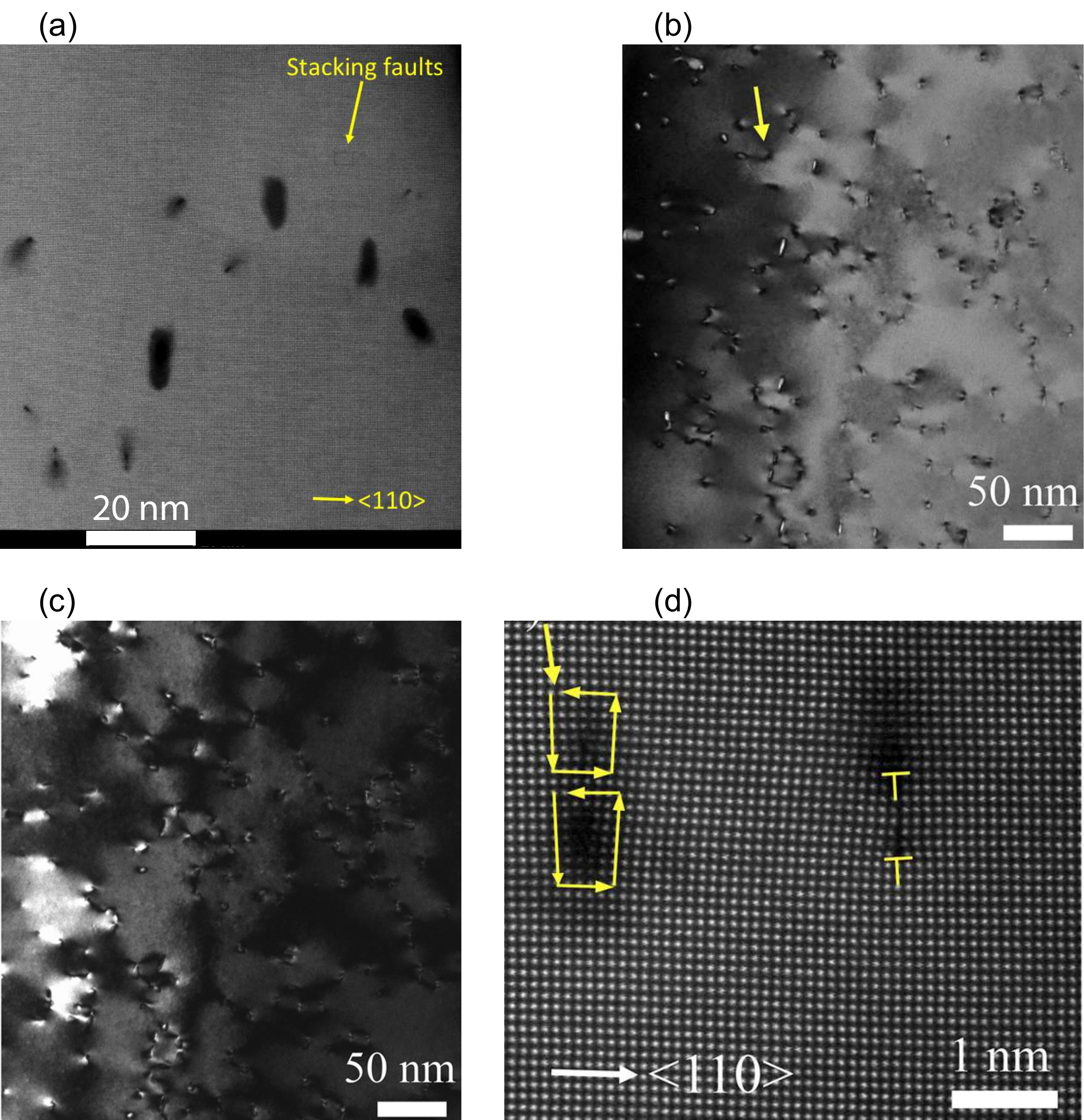}
	\caption{Plan-view STEM images of La-BaSnO$_\text{3}$/BaSnO$_\text{3}$/DyScO$_\text{3}$ film. (a) LM-HAADF STEM image presenting a single loop-shaped stacking fault, designated by a yellow arrow. Bright field (b) and dark field (c) STEM images showing threading dislocations, indicated with dark contrasts in (b), and bright contrasts in (c). The yellow arrow in (b) points to a TD. (d) High resolution HAADF STEM image displaying four edge type threading dislocations, and the Burgers circuit drawn around two of them. The yellow arrow points to the lack of closure of the Burgers circuit. The dislocation edges are represented by the dislocation symbols $\perp$. The images were reproduced and adapted from Ref.~\cite{paik2017adsorption}. Copyright \copyright\, 2017, the Author(s) under a Creative Commons Attribution (CC BY) license. }
	\label{Paik2017a} 
\end{figure}

As compared to the high $\mu_e$ value achieved in single crystals, experimental results have shown that the threading dislocations (TDS), which result from the large lattice mismatch between the film and the substrate, are the main cause of the rather low $\mu_e$ obtained in La:BaSnO$_\text{3}$ thin films \cite{paik2017adsorption,wadekar2014improved,raghavan2016high,mun2013large,yoon2018oxygen,prakash2015hybrid,shiogai2016improvement}. For the epitaxial growth of La:BaSnO$_\text{3}$-based thin films and heterostructures, it is always a challenge to choose the right substrate material with matching lattice parameters. The most challenging parameter to effectively control and prepare high $\mu$ samples is the choice of appropriate substrate~\cite{paik2017adsorption,wadekar2014improved,raghavan2016high,prakash2015hybrid,shiogai2016improvement}. Hence, all the films synthesized using state-of-the-art techniques such as discussed in the previous section experienced a striking number of planar defects such as stacking faults and line defects such as TDs, as revealed by microstructural analyses using transmission electron microscopy (TEM)~\cite{paik2017adsorption,wang2015atomic,raghavan2016high,mun2013large,yoon2018oxygen,prakash2015hybrid,shiogai2016improvement,wadekar2014improved}. 
 
TDs arise from misfit dislocations (MDs), which are created at the film-substrate interface when the mismatch-induced strain  is released \cite{paik2017adsorption,raghavan2016high,yoon2018oxygen,prakash2015hybrid,shiogai2016improvement}. The larger the lattice mismatch,  the more the MDs will be generated at the interface, and the more the TDs will appear in the film.  For the commonly used substrates [see, Fig.~\ref{commercilaLAttice}], the reported percentage of mismatch with La:BaSnO$_\text{3}$ films  are: 3.8\%$-$3.9\% for the TbScO$_\text{3}$~(110) substrates~\cite{lochocki2018controlling}, 3.6\% for the GdScO$_\text{3}$~(110) substrates~\cite{lochocki2018controlling}, 4.2\% for the DyScO$_\text{3}$~(001) substrates~\cite{paik2017adsorption}, 5.0\%$-$5.5\% for the SrTiO$_\text{3}$~(001) substrates \cite{paik2017adsorption,wadekar2014improved,raghavan2016high,lochocki2018controlling,mun2013large,yu2016enhancing,yoon2018oxygen,prakash2015hybrid,lee2016enhanced,shiogai2016improvement}, 3.08\% for the SmScO$_\text{3}$~(110) substrates \cite{wadekar2014improved}, 2.18\%$-$2.3\% for the PrScO$_\text{3}$~(110) substrates \cite{paik2017adsorption,raghavan2016high} and 2.2\% for the MgO substrates~\cite{yoon2018oxygen}.

Figure~\ref{Wang2015} illustrates the different types of structural defects that are often detect in La:BaSnO$_3$ epitaxial films. These are plan-view STEM (scanning transmission electron microscopy) images collected on a La:BaSnO$_3$ film prepared with LaScO$_\text{3}$ buffer layer on  SrTiO$_\text{3}$~(001) substrate using PLD~\cite{wang2015atomic}. Wang \textit{et al.} reported the presence of several types of defects in their films, such as, Ruddlesden-Popper faults, which are a particular type of stacking faults, as well as highly dense shear defects and TDs~\cite{wang2015atomic}. For these particular films, the density of  stacking faults was found to be about $\text{2}\times\text{10}^{\text{11}}$~cm$^{-\text{2}}$. This value is two orders of magnitude larger than the density value of  $\text{3}\times\text{10}^{\text{9}}$~cm$^{-\text{2}}$ for loop-shaped stacking faults observed by Paik \textit{et al.}~\cite{paik2017adsorption}  in their  BaSnO$_3$ buffered film grown by MBE on DyScO$_\text{3}$ substrate. These low density stacking faults are visible in the LM-HAADF STEM micrograph  [see, Fig.~\ref{Paik2017a}\textcolor{blue}{(a)}]. Moreover, from  plan-view STEM data [see, Figs.~\ref{Paik2017a}\textcolor{blue}{(b)} to \ref{Paik2017a}\textcolor{blue}{(d)}], Paik \textit{et al.}  found a density of TDs of $\text{1.2}\times\text{10}^{\text{11}}$~cm$^{-\text{2}}$, which allowed them to conclude that both defect types, stacking faults and TDs, contribute to limit the $\mu$ in their La:BaSnO$_\text{3}$ films. Note that the TDs  are designated in Fig.~\ref{Paik2017a}\textcolor{blue}{(b)}  by the dark contrasts (yellow arrow), in Fig.~\ref{Paik2017a}\textcolor{blue}{(c)} by the bright contrasts, and in  Fig.~\ref{Paik2017a}\textcolor{blue}{(d)} by the dislocation symbols ($\perp$) as well as the lack of closure of the Burgers circuit.

\begin{center} 
\textit{3.2. Boosting electron mobility in La:BaSnO$_3$-based heterostructures by using high temperature grown buffer layers}
\end{center}

For La:BaSnO$_3$ thin films and heterostructures, the main raison for poor $\mu_e$ as compared to the ones measured in single crystals is the large density of defects, and more precisely the TD defects. In the attempt to reduce these TD defects and boost $\mu_e$,  we developed a promising epitatial growth approach for La:BaSnO$_3$-based thin films and heterostructures. This growth approach consists of growing at extremely high temperatures a close lattice matched buffer layer film between the La:BaSnO$_3$ active layer and the substrate. This yields a significant reduction of the TD density and subsequently a boost in $\mu_e$. 

The effectiveness of high-temperature induced TD density reduction had already been established in the case of epitaxial GaAs thin films grown on Si~\cite{takagi1994reduction,deppe1988dislocation}.  In fact, a remarkable reduction of the TD density was observed in high-temperatures annealed GaAs/Si heterostructures, which were then found to form ideal substrate templates for the subsequent growth of device structures such as laser or solar cells~\cite{takagi1994reduction,deppe1987stability}. It was demonstrated that these high thermal stresses exponentially increase the velocity of the glide motion of the  dislocations as well as the concentration of vacancies which enhance their climb motion~\cite{takagi1994reduction,anderson2017theory,choi1977dislocation,yonenaga1987effects}. We clarify here that these vacancies are point defects or interstitials generated by atomic diffusion into the GaAs epilayer upon high thermal treatment. These vacancies may then be trapped by the dislocations, leading either to the direct annihilation of the latter or to an increase in their climb motion~\cite{deppe1988dislocation,takagi1994reduction}.

\begin{figure}[!t]
	\centering 
	\includegraphics[width=1\textwidth]{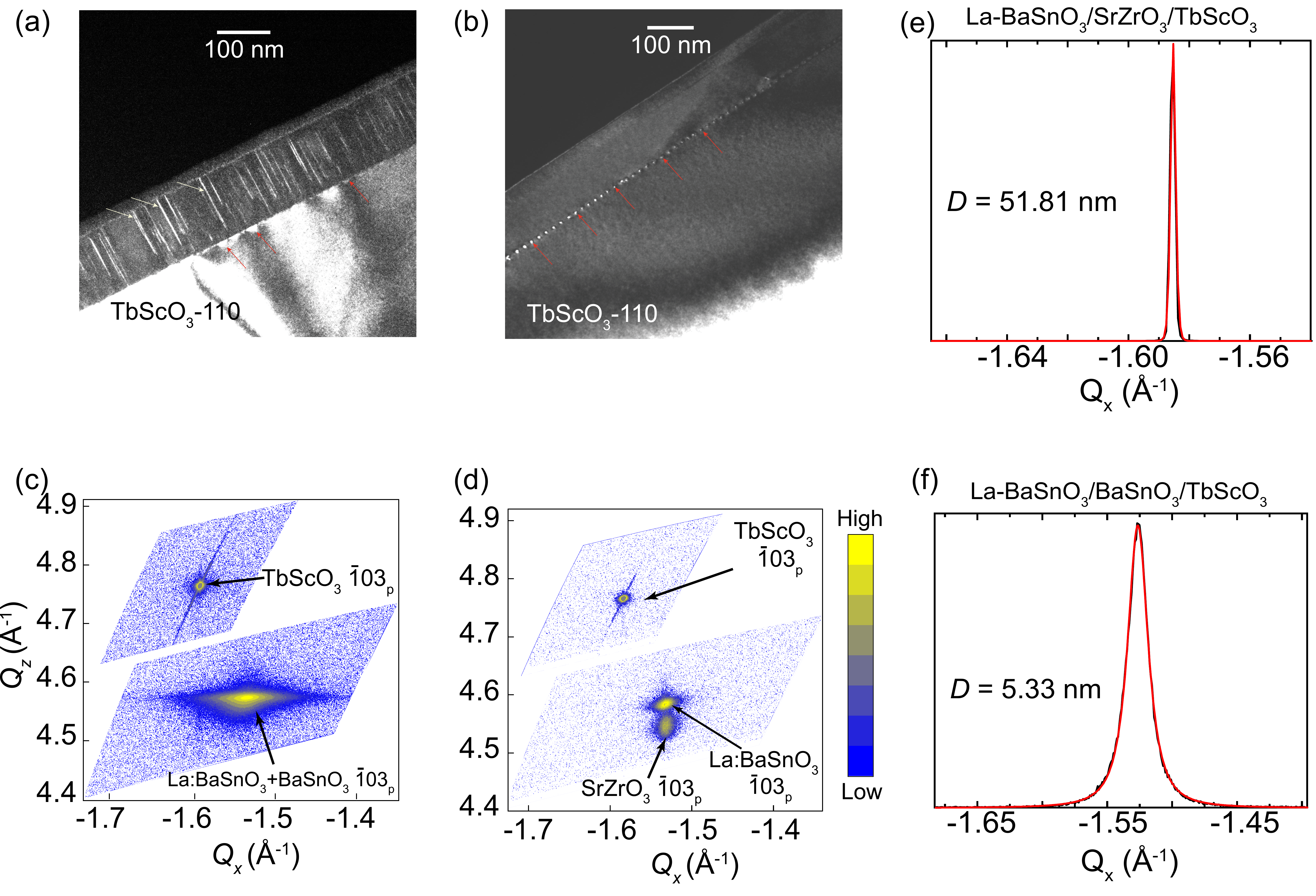}
	\caption{Microstructural and structural data for BaSnO3 and SrZrO3 buffered samples. Weak-beam dark-field TEM images for samples prepared with (a) BaSnO$_3$ and (b) SrZrO$_3$ buffer layers. Misfit dislocations along the interfaces are shown by red arrows. TD are visible in (a), indicated by vertical bright contrasts (white arrows). Only periodic misfit dislocations are visible in (b), suggesting that the theading component of dislocations have annihilated each other Cf Ref.~\cite{nono2019high} of the main text. RSM for (c) BaSnO$_3$ and (d) SrZrO$_3$ buffer layers. Sharper film reflection peaks are observed in the SrZrO$_3$ buffered film, indicating a lower density of TD defects compared to the broader film reflection peak in the BaSnO$_3$ buffered film. The profile contours in (e) and (f) are extracted from the film reflection spots respectively in (c) and (d), and the lateral grain sizes, \textit{D}, obtained after peak function fits of the profiles are indicated. Note the high structural quality of SrZrO$_3$ buffered sample. The images in (a) - (d) are reproduced and adapted from Ref~\cite{nono2019high}. Copyright \copyright\, 2019, the Author(s) under a Creative Commons Attribution (CC BY) license. The images in (e) and (f) are reproduced and adapted from Ref.~\cite{tchiomo2020electronic}. Copyright \copyright\, 2020, the Author(s) under a Creative Commons Attribution (CC BY) license.}
	\label{Figure-09} 
\end{figure}

For the case of La:BaSnO$_3$-based thin films, growing an undoped BaSnO$_3$ buffer layer at high enough substrate temperatures to minimize the density of TD  could be a perfect conceptual approach. However, because of the important volatility of tin oxide at temperatures above 850~$\degree$C, this is not a practical option~\cite{nono2019high}. Alternatively, we turned to adopting the insulating buffer layers of SrZrO$_3$. This buffer layer is temperature stable and could be epitaxially grown at extremely high temperatures due to its low vapor pressure. Also, the SrZrO$_3$ material is the best buffer layer of choice because its lattice parameter is very close to that of La:BaSnO$_3$ material~[see, Fig.~\ref{commercilaLAttice}]~\cite{nono2019high}.  By growing the insulating SrZrO$_3$ buffer layer at an optimal temperature of 1300~$\degree$C, we were able to achieve  $\mu_e^{RT}$ enhancements in active layers of La:BaSnO$_3$ prepared  by using PLD for both active and buffer layers. By replacing BaSnO$_3$ buffer layers grown at 850~$\degree$C with SrZrO$_3$ buffer layers deposited at high-temperature (1300~$\degree$C), the  $\mu_e^{RT}$ was boosted from 117~cm$^\text{2}$~V$^{-\text{1}}$~s$^{-\text{1}}$ for La:BaSnO$_3$/BaSnO$_3$ to 140~cm$^\text{2}$~V$^{-\text{1}}$~s$^{-\text{1}}$ for La:BaSnO$_3$/SrZrO$_3$ heterostructures grown on TbScO$_3$ substrates~ \cite{nono2019high}. This improvement in transport properties was attributed to the exceptional structural properties in these  heterostructures. TEM data showed an impressively low density of TD in the SrZrO$_3$ buffer layers ~[see, Fig.~\ref{Figure-09}\textcolor{blue}{(b)}]. We extracted a value as low as $\text{4.9}\times \text{10}^{\text{9}}$~cm$^{-\text{2}}$, which is two orders of magnitude lower compared to $\text{5}\times \text{10}^{\text{11}}$~cm$^{-\text{2}}$ recorded in La:BaSnO$_3$/BaSnO$_3$ heterostructures [see, Fig.~\ref{Figure-09}\textcolor{blue}{(a)}]. Interestingly, this reduction of the TD desenty is accompanied by a considerable increase in the lateral  size of the crystalline domain (grains), as evidenced by the contour profiles extracted from the reciprocal space maps for the films [see, Fig.~\ref{Figure-09}\textcolor{blue}{(c)} to \ref{Figure-09}\textcolor{blue}{(f)}].

\begin{figure}[!t]
	\centering 
	\includegraphics[width=1\textwidth]{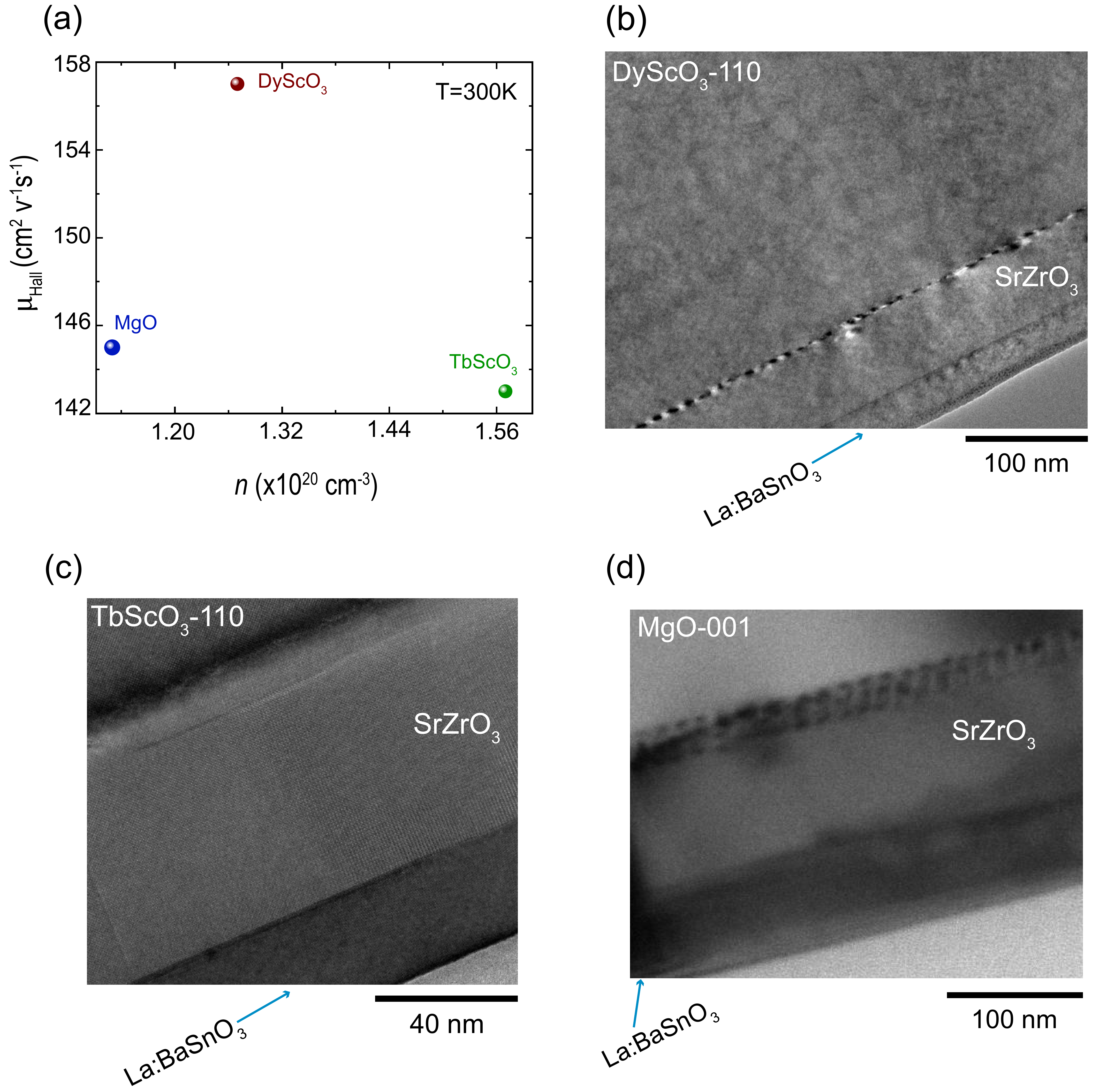}
	\caption{(a) RT electron mobility as a function of the carrier density for La:BaSnO$_3$ (25 nm)/SrZrO$_3$ (100 nm) heterostructures grown on DyScO$_3$ (110), TbScO$_3$ (110), and MgO (001) single crystal substrates. (b)-(d) Bright-field TEM images of the same heterostructures as in (a), showing no evidence of TD defects in the films. Only a network of misfit dislocations is visible along the interface between the films and substrates. Reprinted and adapted with permission from  Ref.~\cite{ngabonziza2023employing}, Copyright \copyright\, 2023 AIP Publishing.}
	\label{Figure-10} 
\end{figure}
Following these improvements, we explored the applicability of this synthesis approach to other oxide substrates including the perovskite DyScO$_3$ and the non-perovskite MgO. For this study, we opted to growing the active La:BaSnO$_3$ layer  by using MBE, since it has been demonstrated that MBE-grown La:BaSnO$_3$ films exhibit better electronic transport characteristics~\cite{paik2017adsorption,wang2019epitaxial,prakash2017wide,raghavan2016high}. Firstly, the SrZrO$_3$ buffer layers were epitaxially grown at a substrate temperature of 1300~$\degree$C with PLD; and then, the deposition of the epitaxial La:BaSnO$_3$ film layers followed using an absorption-controlled MBE process~\cite{ngabonziza2023employing}. Figure~\ref{Figure-10}\textcolor{blue}{(a)} depicts the RT electronic transport properties of these samples. A remarkable improvement of the  $\mu_e^{RT}$ is achieved for the sample prepared on MgO as compared to previously reported values for films grown using MgO substrate [see, Table~\ref{Synopsis-Progress}]. More importantly, a record high  $\mu_e^{RT}$ of 157~cm$^\text{2}$~V$^{-\text{1}}$~s$^{-\text{1}}$ with $\textit{n}=\text{1.27}\times \text{10}^{\text{20}}$~cm$^{-\text{3}}$ is obtained in the sample grown on DyScO$_3$. This is the second-highest $\mu_e^{RT}$ ever achieved in epitaxial La:BaSnO$_3$ films. These outstanding electronic transport properties are associated with the reduction of the TD density in the films due to the high-temperature growth of the SrZrO$_3$ buffer layer. The density of TD in  these heterostructures was very small and it was determined to be as low as $\text{1}\times \text{10}^{\text{10}}$~cm$^{-\text{2}}$ [see, Fig.~\ref{Figure-10}\textcolor{blue}{(b)} to Fig. \ref{Figure-10}\textcolor{blue}{(d)}].
 
The results of these studies evidently indicate that  SrZrO$_3$ epilayer grown at extremely high-temperature on most, if not all oxide substrates constitutes an appropriate template for the subsequent deposition of high $\mu_e$ La:BaSnO$_3$ thin films exhibiting a very low number of TD defects. Moreover, it is noteworthy that achieving high  $\mu_e^{RT}$  in relatively thin samples and with relatively low carrier densities~\cite{nono2019high,ngabonziza2023employing} will enable the fabrication of La:BaSnO$_3$-based FETs on these oxide substrates in which the channels may be fully depleted.

\begin{center}
\textbf{4. Surface and electronic band structures of La:BaSnO$_3$ films and heterostructures}
\end{center}

In addition to the ongoing research activities for enhancing the  $\mu_e^{RT}$ in La:BaSnO$_3$, it is essential to  discuss current research efforts for making a proper correlation between the electronic transport properties and the electronic band structure behaviors of La:BaSnO$_3$ epitaxial films. This will contribute in having a clear understanding of the conduction mechanism in La:BaSnO$_3$ epitaxial films.  The origin of the high $\mu_e^{RT}$ in La:BaSnO$_3$-based materials is attributed to the small $m^*$ of the conducting electrons combined with the largely dispersive Sn~$5s$ orbital in the CB. 

Recently, we reported a systematic investigation of the effect of increasing La doping (or increasing \textit{n}) on the intrinsic spectroscopic properties of La:BaSnO$_3$ films and heterostructures. We explored the core levels, valence band (VB) and CB spectroscopic properties of various La:BaSnO$_3$ films at different La-doping levels. Furthermore, this study provided a better understanding on the chemical purity, the oxidation state, as well as  the surface atomic composition and stoichiometry of La:BaSnO$_3$ material~\cite{nono2022combined}.

\begin{figure}[!t]
	\centering 
	\includegraphics[width=1\textwidth]{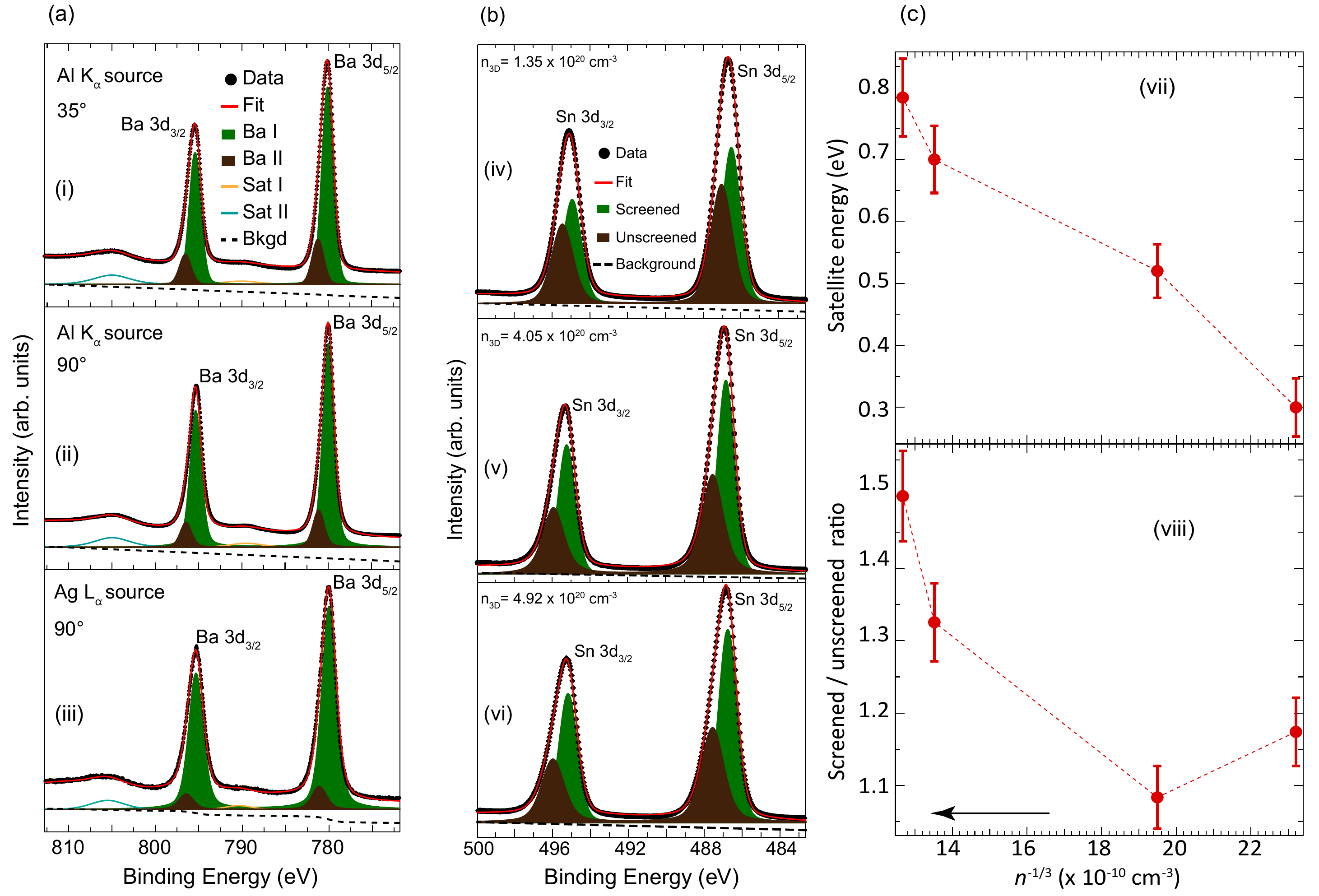}
	\caption{XPS data for the La:BaSnO$_3$ samples. (a) Angle- and energy-dependent XPS spectra of the Ba~$3d$ core level. The spectra in (i) and (ii) were acquired using an Al K$_{\alpha}$ anode at takeoff angles of 35$\degree$ and 90$\degree$, respectively. The spectrum in (iii) was taken using an Ag L$_{\alpha}$ anode at a takeoff angle of 90$\degree$. Two satellite features (cyan and orange peaks) associated with the doublet peak are resolved in the spectra. The peak heights were normalized for clarity. (b) XPS spectra around the Sn 3d regions for three samples of different total	carrier densities as indicated in (iv) - (vi). These data were measured at normal emission with  Al K$_{\alpha}$. (c) Plot of the data exacted from (b), showing the variation of (vii) the satellite energy and (viii) the intensity ratio of the screened	and unscreened peaks as a function of \textit{n}$^{-1/3}$. The dashed lines are guides to the	eye. The black arrow indicates increasing carrier density (\textit{n}). These images are reproduced and adapted from Ref.~\cite{nono2022combined}. Copyright \copyright\, 2022, the Author(s) under a Creative Commons Attribution (CC BY) license.}
	\label{Figure-11} 
\end{figure}

\begin{figure}[!t]
	\centering 
	\includegraphics[width=1\textwidth]{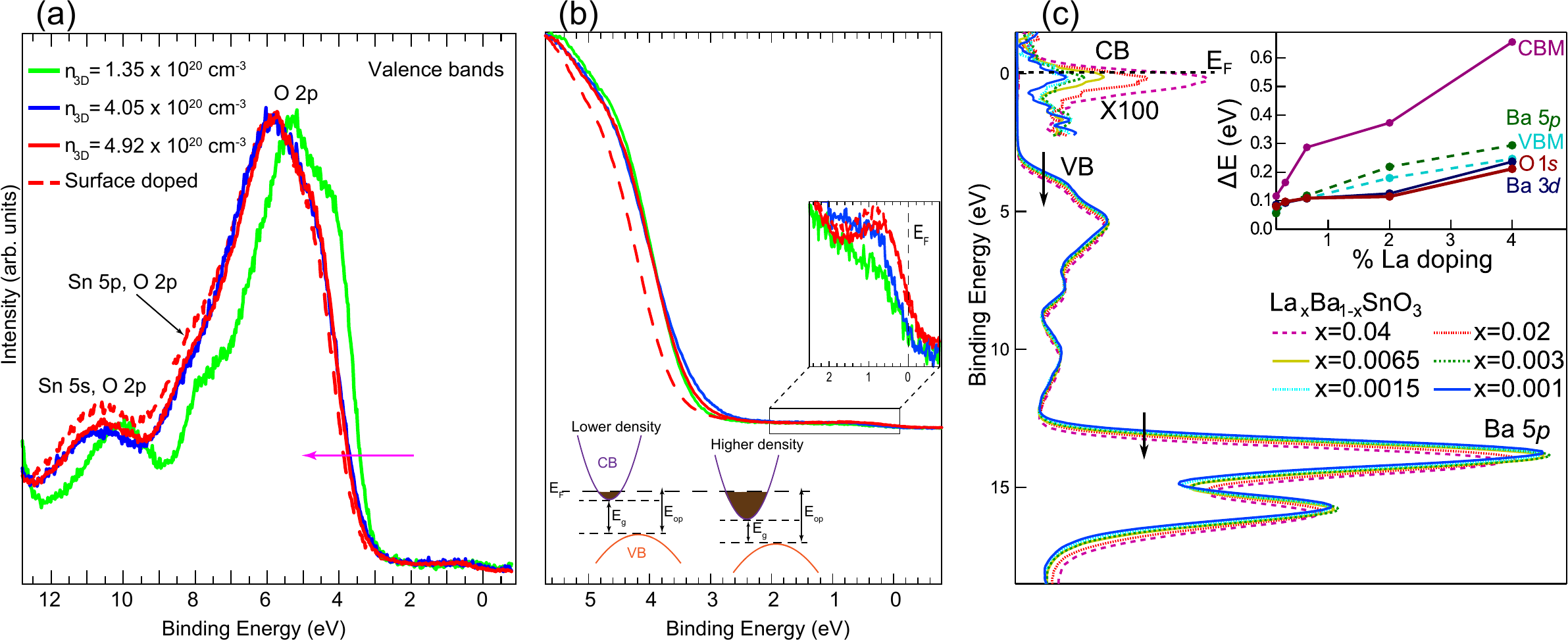}
	\caption{(a) and (b) XPS VB spectra for the same samples shown in Fig.~\ref{Figure-11}. To study the effect of surface absorbed carbonate and hydroxide layers on the states near the VB region, the sample depicted by the red dashed curve (surface doped) was intentionally exposed to contamination after initial	measurements. (a) The magenta arrow points to the direction of the VB shift on increasing \textit{n}. (b) Magnified view of the region around the VB leading edge of the spectra in (a), acquired by adding several scans to achieve an adequate signal to noise ratio. The bottom left inset is a schematic illustration of the Moss–Burstein shift. The top right inset is enlarged spectra for the region around the Fermi level. (c) HAXPES spectra of the core levels, VB and CB regions of several La:BaSnO$_3$ thin films, showing the increasing intensity as well as the shift of the filled CB states with increasing La-doping. The black arrows indicate the direction of the shifts. The inset depicts the energy evolution of the core electrons as well as VB and CB edges with doping. The images in (a) and (b) are reproduced from Ref.~\cite{nono2022combined}. Copyright \copyright\, 2022, the Author(s) under a Creative Commons Attribution (CC BY) license. The image in (c) is reproduced from Ref.~\cite{lebens2016direct}. Copyright \copyright\, 2016, the Author(s) under a Creative Commons Attribution (CC BY) license.}
	\label{Figure-12} 
\end{figure}

Figures~\ref{Figure-11}, Fig.~\ref{Figure-12}\textcolor{blue}{(a)} and Fig.~\ref{Figure-12}\textcolor{blue}{(b)} depict the x-ray photoelectron spectroscopy (XPS) data of the chemical surface characterization of four La:BaSnO$_3$ samples based on fitting the primary core electron line shapes. Fits of the Ba~$3d$ core level are shown in Fig.~\ref{Figure-11}\textcolor{blue}{(a)}. To gain more insights into the doublet peak components structure largely reported for this core level, we  performed the angle-dependent XPS measurements at 35$\degree$ and 90$\degree$ using different XPS sources. We found that Ba~$3d$ spectrum shows an asymmetric line shape, and it was fitted using two symmetric Voigt doublets~\cite{nono2022combined}. The main component labeled Ba~I located at $780.02\pm0.05$~eV binding energy was attributed to lattice barium in the Ba$^{2+}$ state. Whereas the additional doublet peak, labeled Ba~II located at a higher binding energy position of $781.15\pm0.05$~eV was assigned a surface character. This surface character is clearly demonstrated as the ratio of the  Ba~II peak significantly decreases with increasing bulk sensitivity measurements [see, Fig.~\ref{Figure-11}\textcolor{blue}{(a)}]. Increasing the photoemission take-off angle from 35$\degree$ to 90$\degree$, and also increasing the excitation photon energy from 1482.71~eV (Al anode) to 2984.31~eV (Ag anode) provide access to probe states that are located deeper within the sample~\cite{nono2022combined}. The above observations are consistent with the XPS data for Ba~$3d$ core electron measured for powder and epitaxial thin films of BaSnO$_3$~\cite{larramona1989characterization,jaim2017stability}, as well as for epitaxial BaTiO$_3$ thin films\cite{rault2013interface,li2005experimental,li2008characteristics}.

To provide a qualitative and quantitative understanding of the effect of CB filling with increasing carrier density, \textit{n}, in La:BaSnO$_3$ thin films, we analyzed  fits of the Sn~$3d$ core line shapes [see, Fig.~\ref{Figure-11}\textcolor{blue}{(b)}]. As \textit{n} increases,  Sn~$3d$ core spectra present an increasing asymmetry to the high binding energy side of the peak. To take into account the asymmetry in the Sn~$3d$ core spectra, we considered an additional plasmon satellite peak in the analysis. It is known in literature that the asymmetry in core XPS spectra of metallic systems originates from intrinsic plasmon excitations related to the creation of the core hole, which results in an extra component satellite to the main peak~\cite{egdell1999competition,egdell2003screening}. Since the Coulomb potential of the core hole creates a localized trap state (intrinsic plasmon) by pulling out a free electron from the CB~\cite{chazalviel1977final,campagna1975local,korber2010electronic}, the  effect of increasing doping recorded in the line shape of the Sn~$3d$ core electron is most likely associated with screening responses of the free carriers introduced by doping~\cite{egdell1999competition}. Thus, we used two Voigt doublet peaks for the fit of the Sn~$3d$ XPS spectra, assuming that the Koopmans’ state, which is the excited state after the removal of a core electron from the atom, is projected into screened and unscreened final eigenstates~\cite{egdell1999competition}. Therefore, for each spectrum in Fig.~\ref{Figure-11}\textcolor{blue}{(b)}, the main peak is labelled “screened”, and is located at about $486.71\pm0.05$~eV binding energy. The satellite component associated with intrinsic plasmon excitations is labelled “unscreened”~\cite{korber2010electronic}, and is located at about $487.30\pm0.05$~eV~\cite{nono2022combined}. 

Figure~\ref{Figure-11}\textcolor{blue}{(c)} depicts the evolution of both the intensities of the screened and unscreened peaks as well as the energy separation between these components (satellite energy) with increasing \textit{n}. To better visualize the correlation between these parameters, the data are presented as a function of $n^{-1/3}$. The satellite energy and the relative intensity of the screened peak  increase with \textit{n}, while the relative intensity of the unscreened component decreases. These observations are in good agreement with previous studies where a similar plasmon model was employed for the characterization of the Sn~$3d$ and In~$3d$ core photoemission spectra of binary TCO systems such as Sb-doped SnO$_2$, In$_2$O$_3$--ZnO and Sn-doped In$_2$O$_3$~\cite{egdell1999competition,egdell2003screening,chazalviel1977final,korber2010electronic,cox1982free,jia2013direct,christou2000high,Langreth1973TheoryOP}. We highlight that the asymmetry in the Sn~$3d$ core level was also observed in previous reports on La:BaSnO$_3$~\cite{chambers2016band,lebens2016direct} and BaSnO$_3$~\cite{jaim2017stability}  thin films. However, it is striking that a completely different analysis was conducted, and the Sn~$3d$ core XPS spectrum was fitted using three components related to Sn$^{4+}$, Sn$^{2+}$ and Sn metal (Sn$^{0}$)~\cite{jaim2017stability}. This indicates a mixed valence sates for Sn, and points to a non-pristine surface. For our work in Ref.~\cite{nono2022combined}, however, the valence state of Sn was found to be 4+, as demonstrated by the binding energy values of the screened peaks [see, all the Sn 3d spectra in Fig.~\ref{Figure-11}\textcolor{blue}{(b)}]~\cite{morgan1973binding,crist2000handbook}.

The effect of CB filling in the studied La:BaSnO$_3$ samples is also detectable in the VB and CB [see, Fig.~\ref{Figure-12}\textcolor{blue}{(a)} and Fig.~\ref{Figure-12}\textcolor{blue}{(b)}]. This effect is indicated by shifts of the VB leading edge to higher binding energies on  increasing \textit{n}~\cite{nono2022combined}. These shifts are related to the increasing asymmetry in the Sn~$3d$ core levels; and, as reported for other degenerately doped TCOs, these shifts are attributed to the increasing occupancy of the states in the CB~\cite{egdell1999competition,korber2010electronic,jia2013direct}. The spectral features associated with these occupied states are resolved [see, the top right inset in Fig.~\ref{Figure-12}\textcolor{blue}{(b)}]. These are the weak structures  close to E$_F$ also referred to as conduction band minimum (CBM), which terminate in a sharp Fermi edge and whose intensities increase with \textit{n}~\cite{nono2022combined}. We ascribe the observed VB shifts and the associated increase in the intensity of the CB feature to the Moss-Burstein effect, schematically described in the bottom left inset of Fig.~\ref{Figure-12}\textcolor{blue}{(b)}.  The Moss-Burstein effect occurs as the apparent optical band gap is increased following shifts of the onset of interbands absorption to higher energies upon occupation of CB states~\cite{burstein1954anomalous,moss1954interpretation}. These results are consistent with those previously reported for other TCOs~\cite{niedermeier2017electron,lebens2016direct,korber2010electronic,jia2013direct,mudd2014valence}. The analysis presented in Ref.~\cite{nono2022combined} is consistent with the one previously reported for the \textit{ex-situ}  hard x-ray photoemission spectroscopy (HAXPES) data of La:BaSnO$_3$ thin films. This indicates that the core levels, the valence band maximum (VBM), as well as the CBM are effectively altered with increasing \textit{n}~\cite{lebens2016direct} [see, Fig.~\ref{Figure-12}\textcolor{blue}{(c)}].

\begin{figure}[!t]
	\centering 
	\includegraphics[width=1\textwidth]{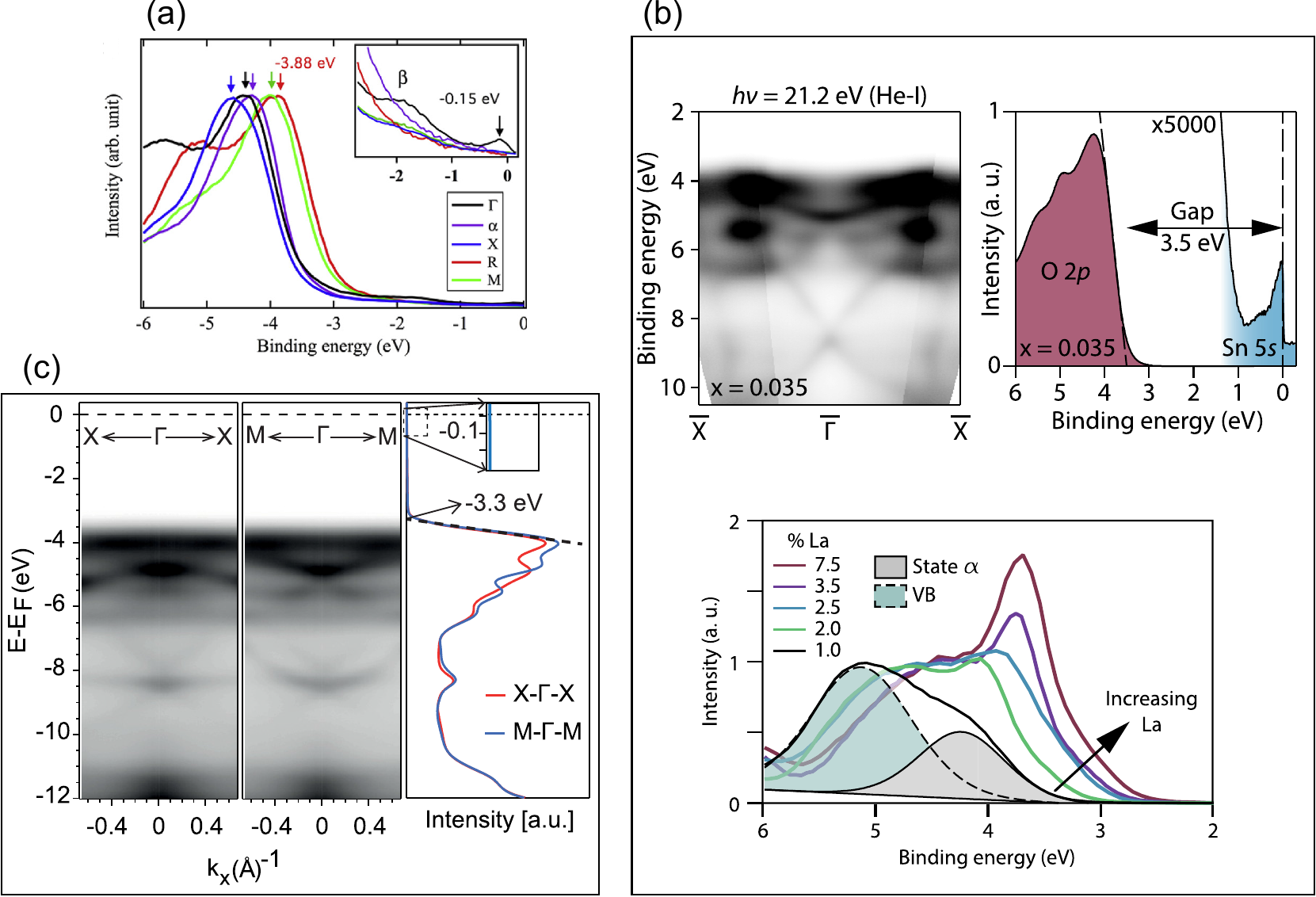}
	\caption{VB structures of (a) and (c) BaSnO$_3$, and (b) La:BaSnO$_3$ thin films measured by ARPES. (a) EDCs at various high symmetry points	indicating the momentum and energy positions of both the VBM and CBM. The inset shows the localized in-gap states $\beta$ as well as occupied states in the CB close to E$_F$ originating from doping induced by native defects such as oxygen vacancies or antisite defects~\cite{joo2017evidence}. (b) Top: ARPES map and its corresponding angle-integrated photoemission spectra. Bottom: EDCs for samples with different La contents showing shifts of the VB to lower energies as well as increase in the intensity of the VB peak $\alpha$  with increasing La concentration. As explained in Ref.~\cite{lochocki2018controlling}, the feature $\alpha$ is attribute to oxygen adsorption. (c) ARPES maps and the corresponding EDCs along high symmetry directions. The inset shows no evidence of in-gap states in the EDCs below E$_F$. The image in (a) is reproduced and adapted with permission from Ref.~\cite{joo2017evidence}. Copyright \copyright\,2016 Elsevier B.V. All rights reserved. The images in (b) reproduced and adapted with permission from Ref.~\cite{lochocki2018controlling}. Copyright \copyright\,2016 AIP Publishing. All rights reserved. The image in (c) Ref.~\cite{soltani20202} is reproduced and adapted with permission from   American Physical Society. Copyright \copyright\,2020 American Physical Society. All rights reserved.}
	\label{Figure-13} 
\end{figure}

\begin{figure}[!t]
	\centering 
	\includegraphics[width=0.9\textwidth]{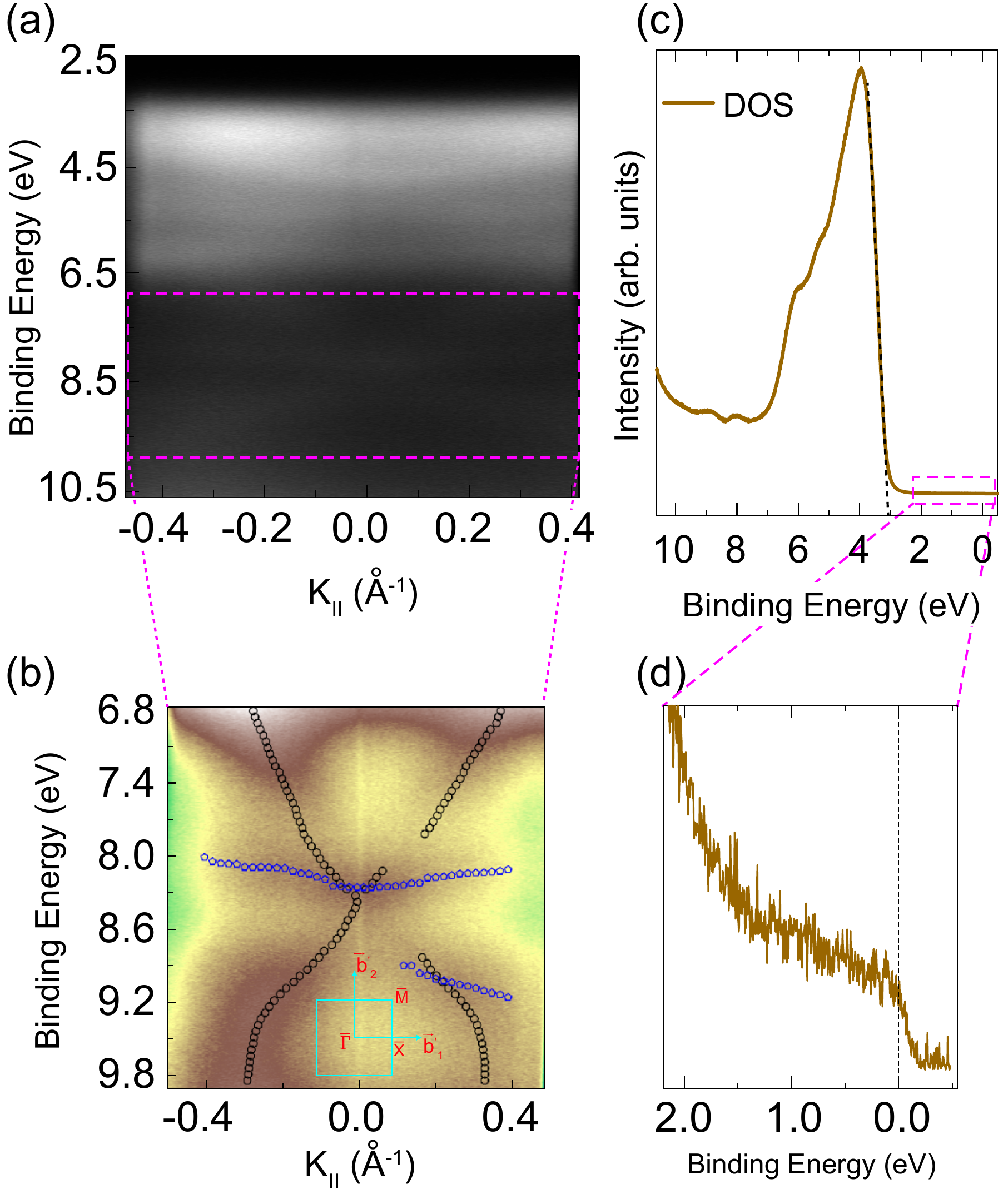}
	\caption{(a) 2D ARPES map for a La:BaSnO$_3$ sample acquired in the $\bar{\Gamma}-\bar{\text{X}}$ direction. (b) ARPES map of the region indicated by the pink dashed rectangle. The inset shows the first Brillouin zone together with the high symmetry points. (c) EDC obtained from the ARPES map in (a). The black dashed	line is a linear extrapolation of the valence band leading edge. (d) Enlargement of the region around the Fermi level in (c), showing the	Fermi–Dirac edge straddling 0~eV. The images are reproduced and adapted from Ref.~\cite{nono2022combined}. Copyright \copyright\, 2022, the Author(s) under a Creative Commons Attribution (CC BY) license.}
	\label{Figure-14} 
\end{figure}

To  date, a few angle resolved photoemission spectroscopy (ARPES) studies  of  the  electronic  band  structure of La:BaSnO$_3$ and BaSnO$_3$ films have been reported~\cite{lochocki2018controlling,joo2017evidence,soltani20202}. ARPES is able to provide a complete characterization of the evolution of the electronic states in the band structure of the La:BaSnO$_3$ samples for different La-doping levels. ARPES measurements give detailed information about the position in momentum space together with the energy position of both the VBM and the CBM, thereby enabling the experimental description of the nature and size of the band gap. For instance, the energy distribution curves (EDCs) at the high symmetry points extracted from \textit{in-situ} ARPES measurements on insulating BaSnO$_3$ films reveal that the VBM is located at the R point with an energetic position of 3.88~eV, and the CBM is situated at the $\Gamma$ point and at -0.15~eV binding energy [see, Fig.~\ref{Figure-13}\textcolor{blue}{(a)}]. This observation suggests that this material has an indirect band gap of 3.7~eV~\cite{joo2017evidence}. Similarly, the \textit{in-situ} ARPES data of La:BaSnO$_3$ films show highly dispersive bands with a VB leading edge located at 3.5~eV, as well as occupied states in the CB below E$_F$ (referenced to 0~eV), yielding an optical band gap of 3.5~eV  [see, upper panel in Fig.~\ref{Figure-13}\textcolor{blue}{(b)}]~\cite{lochocki2018controlling}. The EDCs extracted  from the ARPES spectra near the  $\Gamma$ point reveal that as the percentage of La-doping increases, the VB shifts towards lower binding energies [see, lower panel in Fig.~\ref{Figure-13}\textcolor{blue}{(b)}]. This  opposite trend to the XPS and HAXPES results discussed earlier was proposed to be due to the opposite evolution of surface and bulk chemical potentials, indicating upwards band bending~\cite{lochocki2018controlling}.

In contrast to \textit{in-situ} ARPES experiments which yield highly-resolved bands, \textit{ex-situ} ARPES measurements have proved to be quite challenging in producing well-resolved bands. This is because it requires very particular surface treatment protocols as detailed in Refs.~\cite{tchiomo2020electronic,soltani20202}.  Figure~\ref{Figure-13}\textcolor{blue}{(c)} illustrates the  \textit{ex-situ} ARPES data of BaSnO$_3$ thin films obtained while following a surface preparation recipe that consists of a repeated sequence of  annealing in ultra high vacuum (UHV) and annealing in a tube furnace in varying oxygen pressures~\cite{soltani20202}. Dispersive bands are observed, and from the EDCs extracted from the ARPES data along the high symmetry directions, the VBM is located at 3.3~eV~\cite{soltani20202}. Following the procedure described in Ref.~\cite{NonoTchiomo_2018}, the samples were systematically cleaned in UHV before ARPES measurements. For a  systematic surface treatment procedure which consists of cleaning the La:BaSnO$_3$ samples only in vacuum by annealing in oxygen environment, we were able to resolve highly dispersive bands from \textit{ex-situ} ARPES measurements [see Fig.~\ref{Figure-14}]~\cite{nono2022combined}. These are indicated by the black and blue markers, which correspond to fits of the dispersing bands to peaks in the momentum distribution curves (MDCs) and in the EDCs, respectively [see Fig.~\ref{Figure-14}\textcolor{blue}{(b)}]. The EDC obtained from the ARPES map exhibits features that are similar to those in the VB measured by XPS [see Fig.~\ref{Figure-12}\textcolor{blue}{(a)} and Fig.~\ref{Figure-14}\textcolor{blue}{(c)}], and reveal that the VBM is situated at about 3.1~eV. These ARPES data highlight  the  challenge  of  surface  preparation  for \textit{ex-situ} ARPES measurements  of  epitaxial  La:BaSnO$_3$ films  and  heterostructures. Preferably,  a  portable  vacuum  suitcase  is  ideal  for  long-distance transport  of  epitaxial  La:BaSnO$_3$ films  in  UHV  conditions  to  perform further \textit{ex situ} ARPES analysis at different locations~\cite{PNgabonziza_situ_2018}.

Other than the CBM feature which is not well developed [see, Figs.~\ref{Figure-14}\textcolor{blue}{(d)}], our ARPES data present some similarities with the other experimental reported ARPES spectra~\cite{joo2017evidence,lochocki2018controlling,soltani20202}. It is also in agreement with the calculated dispersions of BaSnO$_3$ and La:BaSnO$_3$. Band structure calculations have predicted that the VB is composed of states such as a mixture of Sn~$5s$ and bonding O~2p orbitals, hybridized Sn 5p and O 2p states, and  bands associated with O 2p bonding or anti-bonding character~\cite{scanlon2013defect,krishnaswamy2017first,themlin1990resonant,farahani2014valence,kover1995high}, which are also resolved in our XPS and ARPES data.

\begin{flushleft}
\textbf{5. Optical properties of epitaxial La:BaSnO$_3$ films}
\end{flushleft}

\begin{figure}[!t]
	\centering 
	\includegraphics[width=1\textwidth]{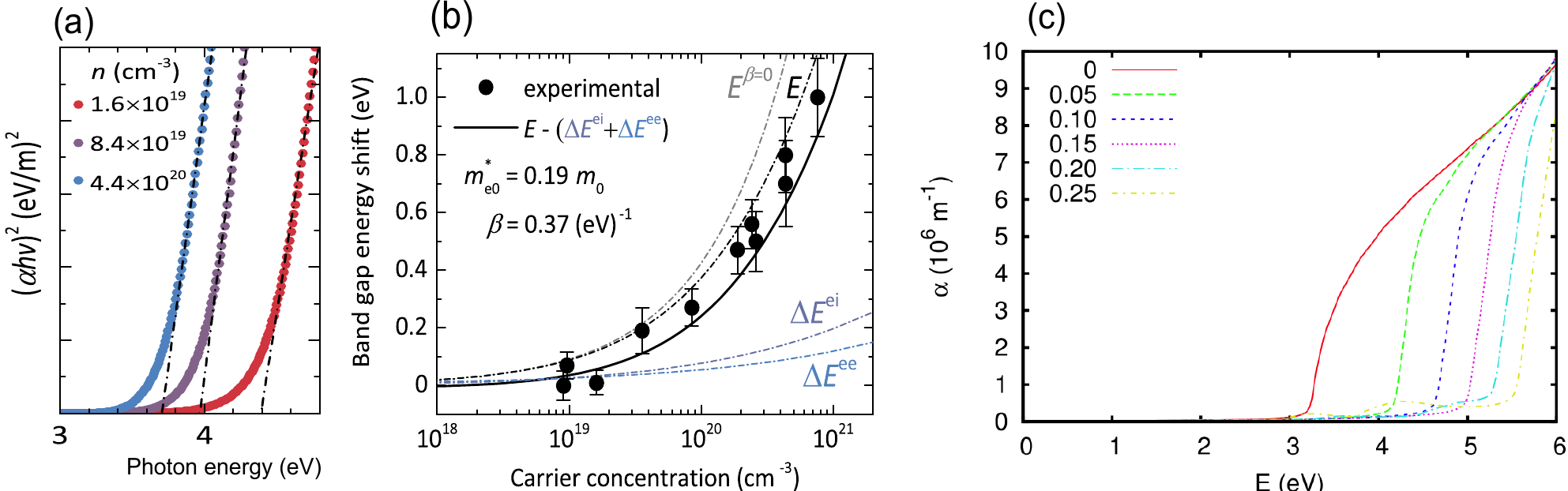}
	\caption{(a) Optical absorption spectra for the  La:BaSnO$_3$/NiO/MgO heterostructures showing shifts of the optical band gap with increasing \textit{n} also shown in (b) (solid black markers).  E$^{\beta=0}$ is the energy dispersion of a parabolic CB, where $\beta$ is an empirical fitting parameter describing the degree of nonparabolicity of the CB. E is the energy dispersion of the CB modelled with the effective mass at the CBM,  $m^{*}_{e0}$,  and $\beta$ as indicated. $\Delta E^{ee}_{g}$ and $\Delta E^{ei}_{g}$ are the band gap narrowing due to electron-electron and electron-impurity interactions, respectively. (c) Calculated absorption spectra for n-type BaSnO$_3$ as a function of doping level in carriers per Sn. The images in (a) and (b) are reproduced and adapted with permission from Ref.~\cite{niedermeier2017electron}. Copyright \copyright\,2017 American Physical Society. All rights reserved. Image in (c) reproduced from Ref.~\cite{li2015tuning}. Copyright \copyright\,2015, Author(s). This article is distributed under a Creative Commons Attribution (CC BY) license.}
	\label{Figure-15} 
\end{figure}

High optical transparency and high DC electrical conductivity constitute the fundamental physical properties that make La:BaSnO$_3$ a functional  material with a wide scope of applications.  The transmittance of undoped BaSnO$_3$ and doped BaSnO$_3$ single crystals (up to 68.7~$\mu$m thick) was found to be 70\% and 20\%, respectively,  in the visible spectral region~\cite{kim2012physical,kim2013indications}. The rather low transmittance in the doped crystals was proposed to be the consequence of their high thicknesses, in combination with the high doping levels. However, it was estimated that the optical transparency would reach as high as 80\% in thin films of about 100~nm~\cite{kim2012physical,kim2013indications}. This prediction was later confirmed for 100~nm thick BaSnO$_3$ thin films sputtered on Si substrates~\cite{du2020bilayer}.

For $520\pm30$~nm thick La:BaSnO$_3$ films prepared on MgO substrates, a transmittance of more than 90\% in the visible was reported~\cite{liu2012composition}, while more than 95\% transparency was recorded for a La:BaSnO$_3$ (300~nm)/SrTiO$_3$ heterostructure, also in the visible region~\cite{wang2007transparent}. Remarkably, the excellent transparency at visible light was reported to be constant in the La:BaSnO$_3$/MgO heterostructures regardless of the doping level. However, for near-infrared radiations (longer wavelengths),  a noticeable fluctuation in the transmission which largely decreases and then somewhat increases again with increasing La-doping content was observed. The decrease of the transmittance at longer wavelength was also reported for the La:BaSnO$_3$/SrTiO$_3$ heterostructure; and also in other TCOs such as La$_{0.5}$Sr$_{0.5}$TiO$_{3+\delta}$~\cite{cho2001optical} and Sn-doped In$_\text{2}$O$_\text{3}$~\cite{ohta2000highly} thin films. This behavior was explained by the full screening of near-infrared lights by free-electrons (Drude-type behavior of the free-carrier excitation in the films), as well as by the shift of the plasma frequency with the charge carrier density, \textit{n}~\cite{liu2012composition,wang2007transparent,cho2001optical,ohta2000highly}. Note that the light is absorbed at wavelengths below $\approx355$~nm, and the sudden sharp increase in the transmittance that follows reflects the development of an optical gap~\cite{liu2012composition}.

\begin{table*}[!b]
	\begin{center}
		\caption{Theoretical prediction of the nature and size of the band gap in BaSnO$_\text{3}$ systems.}
		\label{Table2}
		\setlength{\tabcolsep}{10pt}
		\begin{tabular}{ccc}
			\hline\hline
			Authors & Nature of the band gap  & Size of the band gap (eV)  \\
			\multicolumn{1}{c}{} & \multicolumn{1}{c}{Direct} & 3.46 \\
			\multicolumn{1}{l}{\multirow{-2}{*}{Kang~\textit{et al.} \cite{kang2018first}}} & \multicolumn{1}{c}{Indirect} & 2.98 \\
			\multicolumn{1}{l}{Sallis \textit{et al.}~\cite{sallis2013doped}} & \multicolumn{1}{c}{\multirow{2}{*}{Indirect}} & \multicolumn{1}{c}{\multirow{2}{*}{3.22}}  \\
			\multicolumn{1}{l}{Singh \textit{et al.}~\cite{singh2014strain}} &  &  \\
			\multicolumn{1}{c}{} &  & 3.20 \\
			\multicolumn{1}{l}{\multirow{-2}{*}{Scanlon~\cite{scanlon2013defect}}} & \multicolumn{1}{c}{\multirow{-2}{*}{Indirect}} & 2.49 \\
			\multicolumn{1}{l}{Liu \textit{et al.}~\cite{liu2013origin}} & Direct & 3.00  \\
			\multicolumn{1}{c}{} &  & 2.82 \\
			\multicolumn{1}{l}{\multirow{-2}{*}{Li \textit{et al.}~\cite{li2015tuning}}} & \multicolumn{1}{c}{\multirow{-2}{*}{Indirect}} & 2.66 \\
			\multicolumn{1}{l}{Kim \textit{et al.}~\cite{kim2013hybrid}} & Indirect & 2.48 \\
			\multicolumn{1}{l}{\multirow{1}{*}{Krishnaswamy \textit{et al.}}} & \multicolumn{1}{c}{Direct} & 2.88 \\
			\multicolumn{1}{c}{\multirow{1}{*}{\cite{krishnaswamy2017first,krishnaswamy2016basno3}}} & \multicolumn{1}{c}{Indirect} & 2.40 \\
			\multicolumn{1}{c}{} & \multicolumn{1}{c}{Direct} & 3.96 \\
			\multicolumn{1}{l}{\multirow{-2}{*}{Aggoune \textit{et al.}~\cite{aggoune2022consistent}}} & \multicolumn{1}{c}{Indirect} & 3.50 \\
			\hline\hline
		\end{tabular}
	\end{center} 
\end{table*}

The optical absorption spectra extracted from the transmission data for La:BaSnO$_3$/MgO heterostructures portray the effect of conduction band filling similar to the photoemission spectroscopic observations. The absorption coefficient, $\alpha$, plotted as $\alpha^2$ to determine the size of the band gap~\cite{tauc1966optical} demonstrates that the absorption edge shifts to higher photon energies as the La-doping content increases, indicating an increase in the optical band gap~\cite{liu2012composition}. This increase in the band gap size was explained by the Moss-Burstein shift associated with the degenerate nature of the films. As \textit{n} exceeds the Mott critical density of  the La:BaSnO$_3$ films, the bottom of the CB is filled and more energy will be required to excite the electrons from the VB to the CB, resulting in the widening of the optical band gap~\cite{liu2012composition}.

Optical band gap widening by CB filling was also reported in a study of the dependence of electron $m^*$ on doping levels for La:BaSnO$_3$ (50~nm)/NiO/MgO heterostructures~\cite{niedermeier2017electron}. A direct relationship between the determined optical band gap and \textit{n}  was established. The optical absorption edge was found to shift from 3.7~eV to 4.4~eV with increasing density of conduction electrons [see, Fig.~\ref{Figure-15}\textcolor{blue}{(a)}]. Furthermore, it was reported that the Moss-Burstein  shift associated with increasing \textit{n} is partially compensated by electron-electron and electron-impurity interactions, as the CB edge is shifted to lower energies, resulting in a weak band gap narrowing (0.2~eV narrowing for the highest $\textit{n}=\text{7.6}\times \text{10}^{\text{20}}$~cm$^{-\text{3}}$)~\cite{niedermeier2017electron} [see solid black curve in Fig.~\ref{Figure-15}\textcolor{blue}{(b)}].

Fig.~\ref{Figure-15}\textcolor{blue}{(c)} depicts calculated absorption spectra for n-type BaSnO$_3$ as a function of doping level in carriers per Sn. The calculated optical absorption spectra of BaSnO$_3$ crystals containing different La-doping percentages show an increase in the optical band gap with increasing La-doping~\cite{li2015tuning}. This was suggested to be the result of electrons repulsion within the Sn~$5s$ derived CB due to increase \textit{n}, as well as the shift of the E$_F$ inside the CB as electrons are added~\cite{li2015tuning} [see, Fig.~\ref{Figure-15}\textcolor{blue}{(c)}]. The electrostatic effect raises the energy level of the bands,  consistent with the XPS and HAXPES findings presented above. Table~\ref{Table2} summarizes the reported theoretical predictions on the nature (indirect or direct) and value of the optical band gap  of BaSnO$_3$ systems.

\begin{center}
\textbf{6. High mobility in La:BaSnO$_3$-based thin-film devices}
\end{center} 
\begin{center} 
\textit{6.1. La:BaSnO$_3$-based thin-film field-effect transistors}
\end{center}

La:BaSnO$_3$-based thin film field-effect transistors (TF-FET) exhibiting a high field-effect mobility, $\mu_{FE}$, have the potential for  high-speed and multi-functional transparent devices. Several La:BaSnO$_3$-based TF-FET devices have been designed with the aim to improving not only the $\mu_{FE}$, but also the on-off current ratio, $I_{on}/I_{off}$, and the subthreshold swing, \textit{S}, which is characteristic of the transition speed between  off (low current) and on (high current) states. Hence, it was demonstrated that the structure of the field effect device as well as the selection of the appropriate gate dielectric play a crucial role on the device characteristics. 

\begin{figure}[!t]
	\centering 
	\includegraphics[width=1\textwidth]{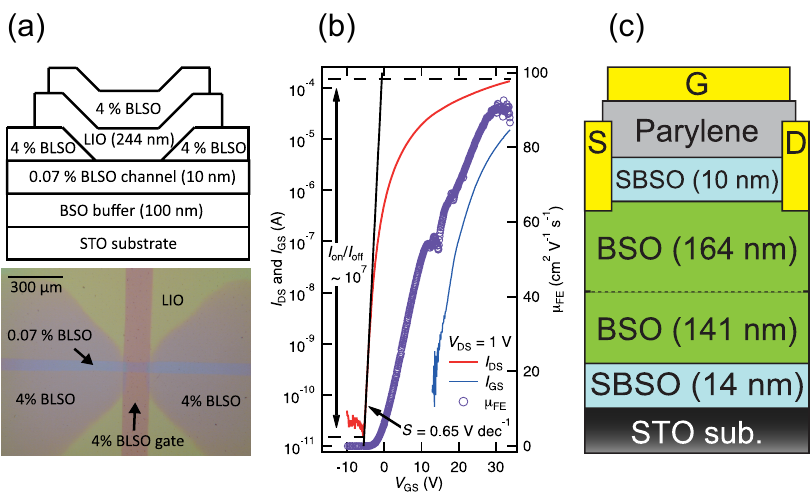}
	\caption{Structures and \textit{I-V} characteristics of La:BaSnO$_3$-based TF-FET devices. (a) Top: schematic of the structure of the all-perovskite FET employing epitaxial thin films of La:BaSnO$_3$ and  LaInO$_3$ as the channel and the gate dielectric, respectively. Bottom: top view image of the same all-perovskite FET obtained by an optical microscope. (b) Transfer characteristics in the linear region for the all-perovskite FET in (a).  (c) Schematic device structure of the FET employing epitaxial thin films of undoped BaSnO$_3$ (BSO) and Sr$_{0.5}$Ba$_{0.5}$SnO$_3$ (SBSO) as channel and barrier layers, respectively, and the organic polymer parylene as a gate dielectric. G, S and D stand for gate, source and drain, respectively. The horizontal dashed line separate the structure regrown on the BSO (141~nm)/SBSO (14~nm)/SrTiO$_3$ (STO) heterostructure to enhance the properties of the device. The images in (a) and (b) are reproduced with permission from Ref.~\cite{kim2015all}. Copyright \copyright\, 2015, Author(s), this article is distributed under a Creative Commons Attribution (CC BY) license. The image in (c) is reproduced and adapted from Ref.~\cite{fujiwara2016high}. Copyright \copyright\,2016, Author(s). This article is distributed under a Creative Commons Attribution (CC BY) license. }
	\label{Figure-16} 
\end{figure}

To exploit the novel and fully versatile properties arising at oxide interfaces~\cite{mannhart2010oxide,zubko2011interface,hwang2012emergent}, La:BaSnO$_3$-based FET devices composed solely of perovskite oxide layers were fabricated~\cite{kim2015all} [see, Fig.~\ref{Figure-16}\textcolor{blue}{(a)}].  In these all-perovskite FET devices, the channel and the gate dielectric were made of epitaxial thin films of La:BaSnO$_3$ and LaInO$_3$, respectively. A record high $\mu_{FE}$ of more than 90~cm$^\text{2}$~V$^{-\text{1}}$~s$^{-\text{1}}$, along with $I_{on}/I_{off}=10^7$ and  $S=0.65$~V~dec$^{-1}$ were achieved [see, Fig.~\ref{Figure-16}\textcolor{blue}{(b)}]. These device performances were attributed to the physical properties of BaSnO$_3$, the high quality of epitaxial LaInO$_3$ film employed as a gate dielectric, and the quality of the LaInO$_3$/La:BaSnO$_3$ heterointerface which could also favor the formation of a two dimensional electron gas (2DEC) due to its polar discontinuity  ~\cite{kim2015all}.

FET device was also fabricated by using a thick insulating BaSnO$_3$ film as the channel, and  an epitaxial film of Sr$_{0.5}$Ba$_{0.5}$SnO$_3$ as a barrier layer [see, Fig.~\ref{Figure-16}\textcolor{blue}{(c)}]. Sr$_{0.5}$Ba$_{0.5}$SnO$_3$ is characterized by a wider band gap, making it suitable for use as a barrier layer between the BaSnO$_3$ channel the gate dielectric. This approach aimed at modulating the band profiles at the heterointerface, since the Sr$_{0.5}$Ba$_{0.5}$SnO$_3$ layer allows conduction band offset due to its larger band gap~\cite{fujiwara2016high}. It was hypothesized that upon applying a positive gate voltage (V$_{GS}$), the Sr$_{0.5}$Ba$_{0.5}$SnO$_3$ barrier acts as a dielectric,  leading to an effective charge accumulation at the Sr$_{0.5}$Ba$_{0.5}$SnO$_3$/BaSnO$_3$ interface. For this heterointerface engineering, an enhanced $\mu_{FE}=52$~cm$^2$~V$^{-1}$~s$^{-1}$ and $I_{on}/I_{off}=10^4$ were reported~\cite{fujiwara2016high}.

By replacing the amorphous Al$_2$O$_3$ gate oxide with the HfO$_2$ one, improved  performance characteristics were reported in the La:BaSnO$_3$-based FET device~\cite{kim2015high,park2014high}. These improvements were attributed to the higher dielectric constant of HfO$_2$, and also to the relatively lower interface trap state at the HfO$_2$/La:BaSnO$_3$ heterointerface (compared with that at the Al$_2$O$_3$/La:BaSnO$_3$ interface). Due to the high dielectric constant of  HfO$_2$, the charge modulation capacity is significantly enhanced in the HfO$_2$/La:BaSnO$_3$ FET~\cite{kim2015high}.

Interface charge traps are the consequence of the high density of TDs that cross the gate dielectric/channel interface and cause extrinsic scatterings. These were demonstrated to have dramatic consequences on the electrical characteristics of the La:BaSnO$_3$-based FET device~\cite{lee2017transparent,kim2015all,fujiwara2016high,kim2015high,park2014high}. It is expected that the performance of La:BaSnO$_3$-based TFT would further improve with the systematic control of the defect level at the gate oxide/channel interface. This could be achieved by either using a lattice-matched substrate, or a more effective buffer layer such as SrZrO$_3$, or a combination of both. The use of a suitable gate dielectric that would favor charge modulation and accumulation at the interface  also offers promising prospects.
 
 \begin{center} 
\textit{6.2. Direction for various La:BaSnO$_3$ thin-film devices and future applications in oxide electronics }
\end{center}

\begin{figure}[!t]
	\centering 
	\includegraphics[width=0.9\textwidth]{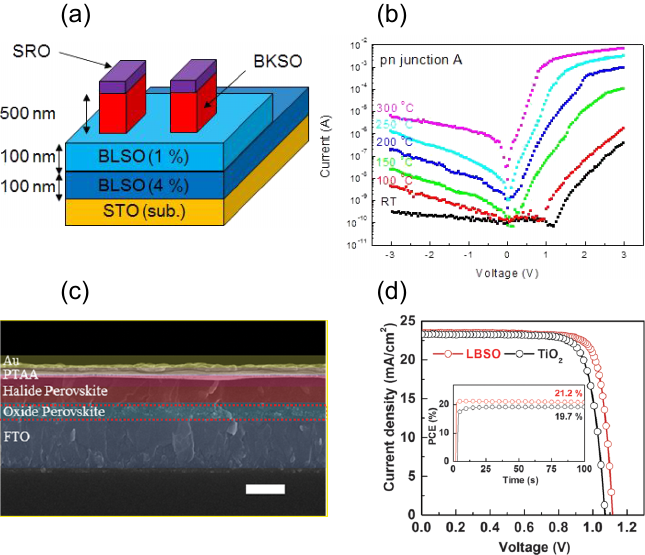}
	\caption{(a) Schematic of the structure of the \textit{p}-K:BaSnO$_3$ (BKSO)/\textit{n}-La:BaSnO$_3$ (BLSO) junction. SrRuO$_3$ (SRO) is the contact used for reliable \textit{p+} contacts to the BKSO films. (b) \textit{I-V} characteristic as a function of temperature for the \textit{pn} junction in (a). (c) Cross-sectional scanning electron microscopy (SEM) image showing the structure of the La:BaSnO$_3$-based PSCs which is composed of: Au counterelectrode, polytriarylamine (PTAA), methylammonium (MA) lead iodine (halide perovskite, MAPbI$_3$) and La:BaSnO$_3$ (oxide perovskite, LBSO) layers on F-doped SO$_2$ (FTO) substrate. The scale bar is 500 nm. (d) Comparison of the \textit{J-V} characteristics for the best-performing PSCs using LBSO and TiO$_2$ (control device) as electron transporting layers. LBSO-based PSC:  open-circuit voltage V$_{oc}=1.12$~V,  short-circuit current density J$_{sc}=23.4$~mA/cm$^2$, and a fill factor $FF=81.3\%$, for an overall PCE of 21.3\%. TiO$_2$-based PSC: V$_{oc}=1.07$~V, J$_{sc}=23.3$~mA/cm$^2$, $FF=78.6\%$ for an overall PCE of 19.6\%~\cite{shin2017colloidally}. The inset shows the stabilized PCEs at a maximum power point (LBSO: 0.96 V; TiO$_2$: 0.91V). The images in (a) and (b) are taken from Ref.~\cite{kim2016thermally}. Copyright \copyright 2016, Author(s). This article is distributed under a Creative Commons Attribution (CC BY) license.
		The image in (c) and (d) are reproduced with permissions from Ref.~\cite{shin2017colloidally}. Copyright \copyright\, 2017, The American Association for the Advancement of Science}
	\label{Figure-17} 
\end{figure}

The potentials of epitaxial La:BaSnO$_3$ films for applications  as diodes and electron transporting layer in solar cell have been explored. For such devices, a \textit{pn} junction is needed, and BaSnO$_3$ should be doped with acceptors on either Ba or Sn sites to exhibit a \textit{p}-type conductivity. \textit{pn} junctions based on single transparent perovskite semiconductor were fabricated by using La:BaSnO$_3$ film as the \textit{n}-type interface, and K-doped BaSnO$_3$ film as the \textit{p}-type semiconductor in which the \textit{p}-type doping was realized by replacing Ba with K~\cite{kim2016thermally} [see, Fig.~\ref{Figure-17}\textcolor{blue}{(a)}]. \textit{I-V} characteristics of these junctions display a typical rectifying behavior, characteristic of a diode-like function [see, Fig.~\ref{Figure-17}\textcolor{blue}{(b)}]. The current level is low at room temperature, and increases with increasing temperature in both the reverse and the forward biases. This was associated with the increasing thermal activation of the \textit{p}-type carriers~\cite{kim2016thermally}. In addition, the junction properties were found to be highly stable upon repeated high-bias and high-temperature  cycling owing to the excellent stability of  oxygen in BaSnO$_3$~\cite{kim2016thermally}. Similar rectifying \textit{I-V} behavior was demonstrated at the \textit{n}-La:BaSnO$_3$/\textit{p}-Si heterojunction, which also displayed photoinduced \textit{I-V} characteristics under violet and green radiations~\cite{luo2014rectifying}.

The integration of La:BaSnO$_3$ as an \textit{n}-type electron transporting layer in solar cells requires La:BaSnO$_3$ films to be deposited onto flexible or even glass substrates. Such process requires temperatures below 300~$\degree$C~\cite{ginley2000transparent,chopra1983transparent,kapur_1999,nakato1995improvement,green1976thin,heilmeier1970liquid,goodman1975liquid,muranoi1978properties,manifacier1981deposition,fortunato2007transparent,shin2017colloidally}. However, the growth of single crystalline La:BaSnO$_3$ film necessitates temperatures above  600~$\degree$C~\cite{nono2019high,tchiomo2020electronic,park2014high,niedermeier2017electron,shiogai2016improvement,shin2016high,prakash2017wide,wang2019epitaxial,paik2017adsorption}, which are too high for the fabrication of optoelectric devices on glass substrates~\cite{shin2017colloidally,huang2015facile}. Very recently, an approach to synthesize single crystalline phase of La:BaSnO$_3$ at temperatures below 300~$\degree$C was developed~\cite{shin2017colloidally}. This technique consisted of preparing a crystalline superoxide-molecular cluster (CSMC) colloidal solution containing well-dispersed CSMC nanoparticles of La:BaSnO$_3$. Hence, compact and uniform layers of La:BaSnO$_3$ were fabricated  by spin-coating the colloidal dispersion of La:BaSnO$_3$ CSMC particles onto F-doped SnO$_2$ glass substrates~\cite{shin2017colloidally}. It was reported that the crystalline phase of the as-prepared La:BaSnO$_3$ films was identical to that of a pure BaSnO$_3$. The La:BaSnO$_3$ layer was subsequently employed as an electrode in perovskite solar cells (PSCs), which exhibited a record power conversion efficiency (PCE) of 21.2\%, as well as an extremely high photostability (less than 10\% change for the PCE) over 1000 hours of solar illumination [see, Fig.~\ref{Figure-17}\textcolor{blue}{(c)} and \ref{Figure-17}\textcolor{blue}{(d)}]. These outstanding performances were attributed to the intrinsic properties of La:BaSnO$_3$, including its high electron density which might induce a large potential built-in at the heterojunction with the halide perovskite layer~\cite{shin2017colloidally,myung2018doped}.

As this review paper intends to provide a comprehensive study of the physical characteristics of epitaxial La:BaSnO$_3$ thin films, as well as an overview of potential device applications implementing these properties, BaSnO$_3$-based applications will not be addressed here in detail. The interested reader can consult available references on the topics of BaSnO$_{3-\delta}$- and BaSnO$_3$-based FET and \textit{pn} junctions~\cite{lee2017transparent,luo2017magnetically,du2020bilayer,lee2017transparent1,zhang2018interface,wang2020electronic} and also available references on the topic of BaSnO$_{3}$-based solar cells and photoconductors~\cite{li2010fabrication,shin2013improved,kim2013basno3,zhu2016mesoporous,park2016photoconductivity}.
\begin{center} 
\textit{6.3. Fundamental realization of 2DEG with high $\mu_e$ in La:BaSnO$_3$-based heterostructures}
\end{center} 

The polar discontinuity between LaInO$_3$ and La:BaSnO$_3$ layers was suggested to result in the formation of a two-dimensional electron gas (2DEG) at their heterointerface~\cite{kim2015all}. Also, the LaInO$_3$/La:BaSnO$_3$ interface forms an excellent basis for carrier accumulation due to the lattice matching between the two perovskite oxides, which helps in reducing interfacial scattering associated with structural defects at the interface~\cite{kim2015all}. 

\begin{figure}[!t]
	\centering 
	\includegraphics[width=1\textwidth]{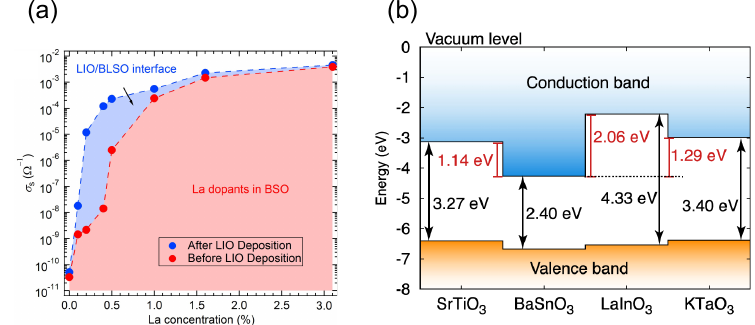}
	\caption{(a) Evidence of 2DEG sheet conductance enhancement in LaInO$_3$ (LIO)/La:BaSnO$_3$ (BLSO) interfaces with increasing La-doping, compared with the conductance at the bare BLSO surfaces. (b) Natural band alignment between BaSnO$_3$, SrTiO3$_3$, LaInO$_3$, and KTaO$_3$ calculated from first principles. The BaSnO$_3$ band structure was referenced to the vacuum level based on a calculation for an SnO$_2$-terminated surface. The image in (a) is taken from Ref.~\cite{kim2016conducting}. Copyright \copyright 2016, Author(s). This article is distributed under a Creative Commons Attribution (CC BY) license. The image in (b) is reproduced with permission from Ref.~\cite{krishnaswamy2016basno3}. Copyright \copyright\,2016 AIP Publishing. All rights reserved.}
	\label{Figure-18} 
\end{figure}

Highly conducting 2DEG (with conductance as large as a factor of $10^4$) was reported at the LaInO$_3$/La:BaSnO$_3$ polar interface~\cite{kim2016conducting}. As the concentration of La in the BaSnO$_3$ layer increases, the conductivity $\sigma$, the carrier density  \textit{n} and the mobility $\mu_e$   of this 2DEG are enhanced  [see, Fig.~\ref{Figure-18}\textcolor{blue}{(a)}]. A $\mu_e^{RT}$ as high as $80~\text{cm}^2~\text{V}^{-1}~\text{s}^{-1}$ and \textit{n} as high as $2\times 10^{14}$~cm$^{-2}$ were then reported~\cite{kim2016conducting}. After ruling out the possibility that oxygen vacancy and La diffusion between La:BaSnO$_3$ and LaInO$_3$ layers could be involved in these electrical properties enhancement, it was demonstrated that electronic reconstruction at the LaInO$_3$/La:BaSnO$_3$ polar interface is the main reason for the development of enhanced interfacial 2DEG conductivity~\cite{kim2016conducting}. This experimentally observed 2DEG was explained using band calculations based on self-consistent 1D Poisson-Schr{\"o}dinger equations, considering the interfacial polarization in LaInO$_3$~\cite{kim2019interface,shin2020remote}. In addition to the 2DEG dependence on the La-doping content in the BaSnO$_3$ layer,  the thickness of the LaInO$_3$ layer also strongly influences the 2DEG carrier density~\cite{kim2019interface,shin2020remote,aggoune2021tuning}.

As the TDs are the most important source of mobility limitation in BaSnO$_3$ system, engineering an interface that exhibits a low density of TD is then the key for realizing a high  $\mu_e^{RT}$ 2DEG in BaSnO$_3$-based system~\cite{eom2022oxide}. A 2DEG with a large $\mu_e^{RT}=60~\text{cm}^2~\text{V}^{-1}~\text{s}^{-1}$ and $\textit{n}=1.7\times 10^{13}$~cm$^{-2}$  was realized at the polar LaScO$_3$/BaSnO$_3$ heterointerface~\cite{eom2022oxide}. The high RT 2DEG $\mu$ was achieved upon engineering the LaScO$_3$/BaSnO$_3$ interface so as to obtain lattice-matching, leading to a reduced TD density and a subsequent enhancement of the electrical characteristics of the conducting channel. This suggests that an even higher $\mu_e^{RT}$ for interfacial 2DEG could be achieved if La:BaSnO$_3$ thin film is used as a host instead of BaSnO$_3$ film~\cite{eom2022oxide}. This is because the Coulomb scattering of conduction electrons can be significantly reduced as the La ions could concentrate inside the TD cores (characterized with abundant dangling bonds~\cite{dexter1952effects,ng1998role,podor1966electron,weimann1998scattering}) and form anti-site defects that will screen the potential of positively charged oxygen vacancies known to occupy dislocation cores in perovskite oxides~\cite{lubk2013electromechanical,marrocchelli2015dislocations,hirel2016electric}.

To produce interfacial 2DEG in BaSnO$_3$-based bilayer heterostructures, a proper band structure alignment is necessary, considering the conduction band offset (CBO) of the BaSnO$_3$ host system with the other materials [see, Fig.~\ref{Figure-18}\textcolor{blue}{(b)}]~\cite{lee2017transparent,eom2022oxide,kim2016conducting,kim2015all,krishnaswamy2016basno3}. A study addressing the quantitative confinement of the 2DEG at BaSnO$_3$-based heterointerface has been conducted, and the confined 2DEG density as a function of the CBO of BaSnO$_3$ with any polar or non-polar material used as barrier was proposed~\cite{krishnaswamy2016basno3}. The 2DEG density increases with increasing CBO, and for a CBO of 3.2~eV, 2DEG in the order of $10^{14}$~cm$^{-2}$ could be confined~\cite{krishnaswamy2016basno3}. Hence, large enough CBO along with perfect lattice-matching  are required to achieve  highly mobile 2DEG in BaSnO$_3$-based bilayer heterointerface~\cite{eom2022oxide,kim2015all,krishnaswamy2016basno3}.
\begin{center}
\textbf{7. Conclusion and outlook }
\end{center}

In summary, recent progress in epitaxial film growth, and the investigation of the electronic transport properties and the electronic band structure features of La:BaSnO$_3$ thin films are presented. Recent successful efforts to fabricate high mobility La:BaSnO$_3$-based thin-film devices were also discussed.  By using the combination of PLD and MBE growth techniques coupled with a CO$_2$ laser substrate heating technology, we demonstrated an approach to use buffer layers grown at  very  high  temperatures  for  the  reduction  of  TDs  and then   improve the  room temperature electron mobility in epitaxial La:BaSnO$_3$ heterostructures grown on several oxide  substrates. The density of TDs are very low in these films with an upper limit well below 1.0$\times$10$^{10}$cm$^2$ for all films prepared on various oxide substrates, thus verifying the efficacy of our synthesis approach. This methods provides an effective approach for the growth of high mobility La:BaSnO$_3$ epitaxial films on most, if not all, oxide substrates that present large compressive or tensile lattice mis-matches to La:BaSnO$_3$, which is an essential step in tackling the challenges caused by the lack of commercially available substrates with lattice parameters matching the BaSnO$_3$  unit cell. Furthermore, these high-quality thin film heterostructures offer the flexibility for performing systematic investigation of the surface and electronic band structures of La:BaSnO$_3$ films, as well as the optical properties of these epitaxial films. Achieving high $\mu_e^{RT}$ at low thickness and relatively low carrier concentrations in these La:BaSnO$_3$-based heterostructures provides an opportunity to fabricate La:BaSnO$_3$-based FETs on various oxide substrates in which channels may be fully depleted. Furthermore, we review recent direction for the fabrication of various La:BaSnO$_3$ thin-film devices and their potential future applications for oxide electronics. Lasty, we present very recent progress in the fundamental realization of 2D electron gases with high electron mobility in La:BaSnO$_3$-based heterostructures.

As outlook, with this promising progress in the epitaxial growth of La:BaSnO$_3$ heterostructures, it is only a matter of time before the $\mu_e^{RT}$ in buffered epitaxial La:BaSnO$_3$ thin films reproducibly reach the $\mu_e^{RT}$   values reported in single crystals. This will pave a way towards the exploration of the physics of La:BaSnO$_3$-based devices and for their application in the next-generation of electronic and optoelectronic devices.
\begin{center}
\textbf{ Acknowledgments}
\end{center}

P. Ngabonziza. acknowledges startup funding from the College of Science and the Department of Physics \& Astronomy at Louisiana State University. 
\newpage

\bibliography{references_01}

\end{document}